\documentclass[twocolumn,english]{revtex4-1}
\usepackage[T1]{fontenc}
\usepackage{color}
\usepackage{amsmath}
\usepackage{amssymb}
\usepackage{graphicx}
\usepackage{esint}
\usepackage{lineno}

\definecolor{ablue}{rgb}{0.1,0.35,0.75}
\definecolor{agreen}{rgb}{0,0.6,0.4}

\usepackage[normalem]{ulem}

\makeatletter
\@ifundefined{textcolor}{}
{%
 \definecolor{BLACK}{gray}{0}
 \definecolor{WHITE}{gray}{1}
 \definecolor{RED}{rgb}{1,0,0}
 \definecolor{GREEN}{rgb}{0,1,0}
 \definecolor{BLUE}{rgb}{0,0,1}
 \definecolor{CYAN}{cmyk}{1,0,0,0}
 \definecolor{MAGENTA}{cmyk}{0,1,0,0}
 \definecolor{YELLOW}{cmyk}{0,0,1,0}
}

\renewcommand{\fnum@figure}{\textbf{Figure~\thefigure}}
\usepackage{units}
\usepackage{lipsum}
\usepackage{babel}

\renewcommand{\vec}[1]{{\mathbf{#1}}}
\newcommand{\mi}{\mathrm{i}}

\makeatother

\usepackage{babel}
\begin{document}

\title{Hybridized intervalley moir\'e excitons and flat bands in twisted WSe$_2$ bilayers}

\author{Samuel Brem$^1$}
\author{Kai-Qiang Lin$^2$}
\author{Roland Gillen$^3$}
\author{Jonas M. Bauer$^2$}
\author{Janina Maultzsch$^3$}
\author{John M. Lupton$^2$}
\author{Ermin Malic$^1$}
\affiliation{$^1$Chalmers University of Technology, Department of Physics, Gothenburg, Sweden}
\affiliation{$^2$University of Regensburg, Institute of Experimental and Applied Physics, Regensburg, Germany}
\affiliation{$^3$Friedrich-Alexander University Erlangen-Nuernberg, Institute of Condensed Matter Physics, Germany}
\begin{abstract}
The large surface-to-volume ratio in atomically thin 2D materials allows to efficiently tune their properties through modifications of their environment. Artificial stacking of two monolayers into a bilayer leads to an overlap of layer-localized wave functions giving rise to a twist angle-dependent hybridization of excitonic states. In this joint theory-experiment study, we demonstrate the impact of interlayer hybridization on bright and momentum-dark excitons in twisted WSe$_2$ bilayers. In particular, we show that the strong hybridization of electrons at the $\Lambda$ point leads to a drastic redshift of the momentum-dark K-$\Lambda$ exciton, accompanied by the emergence of flat moir\'e exciton bands at small twist angles. We directly compare theoretically predicted and experimentally measured optical spectra allowing us to identify photoluminescence signals stemming from phonon-assisted recombination of layer-hybridized dark excitons. Moreover, we predict the emergence of additional spectral features resulting from the moir\'e potential of the twisted bilayer lattice.    
\end{abstract}
\maketitle

Manufacturing of artificially stacked multilayer materials has recently become technologically feasible \cite{geim2013van, kunstmann2018momentum, zhang2018moire, merkl2019ultrafast}. In particular, encapsulation techniques using hexagonal boron nitride have greatly improved the homogeneity of monolayer material properties \cite{dean2010boron,raja2019dielectric}.  Apart from the possibility to further miniaturize existing semiconductor technologies, the large surface-to-volume ratio in monolayers of transition metal dichalcogenides (TMDs) additionally allows to externally tailor the materials' properties in an unprecedented scope. More specifically, the stacking of two monolayers and the resulting overlap of electronic wave functions gives rise to a hybridization of the corresponding quantum states \cite{fang2014strong,coy2015direct,alexeev2019resonantly}. In homo bilayers, where the electronic bands of the composing monolayers are energetically degenerate, the interlayer hybridization becomes dominant and significantly modifies the bilayer eigenstates. Recent studies have shown that the hybridization and thus the bilayer band structure can be externally tuned by controlling the stacking angle \cite{van2014tailoring,yeh2016direct}. Moreover, moir\'e patterns created through the misaligned monolayer lattices allow to create tailored arrays of trapping potentials \cite{yu2017moire,wu2018theory, rivera2018interlayer,tran2019evidence,seyler2019signatures} and potentially give rise to flat bands and superconductivity \cite{wu2018hubbard,wang2019magic,an2019interaction,zhang2019flat} similar to the case of twisted bilayer graphene.

Studies on interlayer hybrid moir\'e excitons have so far mostly focused on the optically active intravalley excitons located at the K point \cite{jin2019observation,tran2019evidence,seyler2019signatures,alexeev2019resonantly}. While this restriction is often sufficient in monolayers due to their direct band gap at the K point, the conduction band minimum, e.g. in tungsten-based homo bilayers, is located rather at the $\Lambda$ point and the associated optically dark intervalley K-$\Lambda$ excitons represent the energetically lowest and therefore dominant exciton species \cite{lindlau2018role,deilmann2019finite,berghauser2018mapping}. In this work, we combine density matrix formalism, ab initio calculations and optical experiments to investigate layer-hybridized intervalley excitons in twisted bilayer WSe$_2$ (tWSe$_2$). We find that hybridization and resulting moir\'e effects are small at the K point giving rise to either intra- or interlayer excitons. In contrast, electrons at the $\Lambda$ point are strongly delocalized across both layers, resulting in a drastic redshift of the K-$\Lambda$ exciton and in the emergence of flat moir\'e bands at small twist angles.  In good agreement between theory and experiment, we find clear twist-angle dependent photoluminescence signatures of the dark K-$\Lambda$ excitons in tWSe$_2$.

\begin{figure}
\includegraphics[width=\columnwidth]{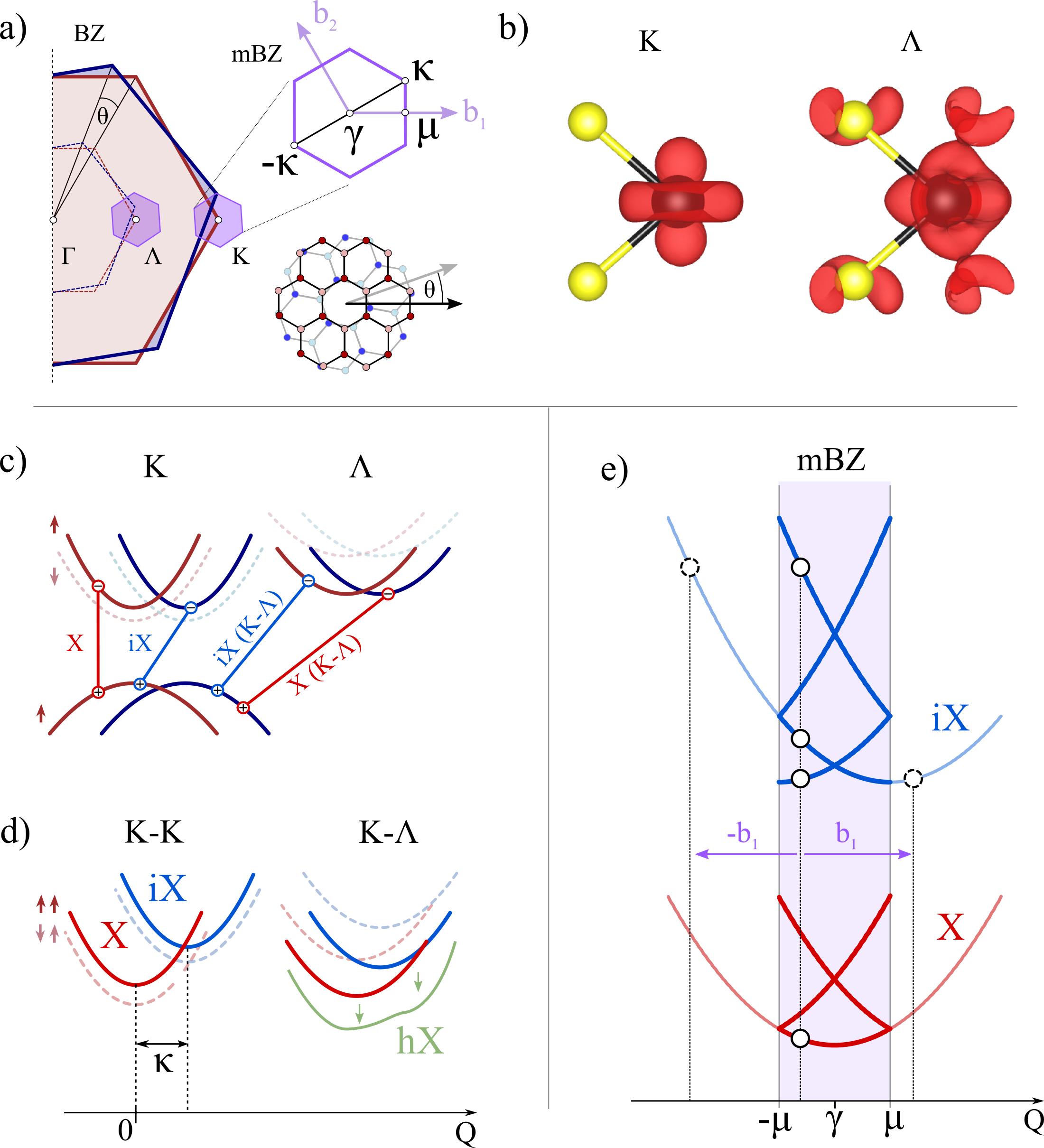}
\caption{Exciton hybridization in twisted bilayer TMDs. (a) High-symmetry points in the Brillouin zone (BZ) and mini-Brillouin zone (mBZ) at a specific twist angle $\theta$. (b) Partial charge density for conduction band electrons at the K and $\Lambda$ point obtained with DFT calculations. (c) Schematic electronic bandstructure at the K and $\Lambda$ valley of the two twisted layers (red and blue, respectively) as well as possible intra- (X) and interlayer excitons (iX). (d) Schematic exciton center-of-mass dispersion without hybridization as well as hybridized K-$\Lambda$ state hX (green). Intra- and interlayer K-K' dispersions are shifted by the momentum $\kappa\sim\theta$.  (e) The periodic mixing of discrete momenta at the K point can be described by the mixing of different subbands with the same momentum in a zone folding scheme (cf. open circles).}
\label{fig:concept} 
\end{figure}

\textbf{Excitonic bandstructure.}
In this work, we focus on the low energy excitations close to the band edges at different high symmetry points of the hexagonal Brillouin zone (BZ) in TMDs, cf. Fig. \ref{fig:concept}(a). Here, minima of the conduction band are located at $K,\Lambda,\Lambda',K'$, while valence band maxima can be found at $K,\Gamma,K'$. When the two layers are twisted by an angle $\theta$, the two monolayer BZs rotate accordingly (Fig. \ref{fig:concept}(a)). The overlap of electronic wave functions of the two layers gives rise to a hybridization of electronic states. Hereby, only states with the same quasi-momentum can hybridize, so that the twist angle is the key parameter determining which parts of the two band structures mix. 

The interlayer interaction strength is determined by the wave function overlap of the two adjacent layers and is therefore strongly valley-dependent. The conduction band wave function e.g.  at the $\Lambda$ point has a significant contribution at selenium atoms giving rise to a strong overlap, cf. Fig. \ref{fig:concept}(b). In contrast, at the K point the wave function is mostly composed of d orbitals localized at tungsten atoms, so that the K point electrons are well protected from the environment \cite{cappelluti2013tight,roldan2014electronic}. We apply density functional theory to calculate parameters for the interlayer interaction strength and use standard literature parameters for the electronic bandstructure in single monolayers (effective masses, valley separations). These input parameters are used to set up a realistic exciton model in the vicinity of band extrema within the density matrix formalism. 
Fig. \ref{fig:concept}(c) schematically illustrates valence and conduction band of the two monolayers (red and blue) for the most important K and $\Lambda$ valley. As a result of the strong Coulomb interaction in 2D systems, electrons and holes in TMDs are strongly bound into excitons \cite{he2014tightly}. A monolayer can host intravalley excitons at the K point (denoted by X) as well as intervalley excitons \cite{wang2018colloquium,mueller2018exciton} such as X(K-$\Lambda$), where electron and hole are located at the $\Lambda$ and the K valley, respectively. In a bilayer system, we additionally find interlayer excitons (denoted by iX) with electrons and holes located in different layers \cite{kunstmann2018momentum,rivera2018interlayer,merkl2019ultrafast,ovesen2019interlayer}. Although spatially separated, they still have binding energies in the range of  100meV \cite{merkl2019ultrafast, ovesen2019interlayer} and can also be momentum-indirect. In order to account for the strong Coulomb interaction in TMDs, we transform the bilayer Hamiltonian into an exciton basis, giving rise to a modified energy landscape in terms of intra- and interlayer-type excitons.

Fig. \ref{fig:concept}(d) shows schematically the exciton  dispersion in a bilayer without  (red and blue) and with accounting for hybridization effects (green). In the exciton basis, the hybridization of electronic states corresponds to a mixing of intra- and interlayer excitons. The energy of the hybrid exciton hX(K-$\Lambda$) can be approximated by the well-known avoided crossing formula
\begin{eqnarray}\label{eq:anti-crossing}
E^{\text{hX},\pm}_{\mathbf{Q}}=\dfrac{1}{2}(E^\text{iX}_{\mathbf{Q}} + E^\text{X}_{\mathbf{Q}}) \pm  \dfrac{1}{2} \sqrt{ \Delta E_{\mathbf{Q}}^2 + 4 |T|^2 },
\end{eqnarray}
with $\Delta E_{\mathbf{Q}}=E^\text{iX}_{\mathbf{Q}} - E^\text{X}_{\mathbf{Q}}$.  The interlayer interaction strength $T$  contains the overlap of electronic states weighted by the in-plane excitonic wave functions, cf. supplementary material. Note that in Eq. \ref{eq:anti-crossing} only excitons with the same center-of-mass momentum can hybridize. When considering interlayer hopping of electrons/holes at the K point this momentum conservation is modified. Since there are three equivalent K points within the first BZ, a K point electron in one layer mixes with three momenta of the other layer.
 The mixing of discrete momenta in different layers can be interpreted in real space as a quasi-free  electron interacting with a periodic potential created by the moir\'e pattern. Consequently, this super lattice gives rise to a zone folding into a mini-Brillouin zone (mBZ) [Fig. \ref{fig:concept}(a)] and the emergence of a series of subbands illustrated in Fig. \ref{fig:concept}(e). Within the mBZ the hybridization of bands only occurs at a fixed momentum, however the mixing can involve different moir\'e subbands \cite{ruiz2019interlayer}. In contrast to the K point, high-symmetry points deep within the BZ (such as $\Lambda$ and $\Gamma$) only have equivalent states outside of the first BZ. Since interlayer hopping strongly decreases for large momenta, this additional mixing is negligible \cite{wang2017interlayer}, so that the dispersion at the $\Lambda$ and $\Gamma$ valley does not split into subbands. 

 Based on ab initio parameters \cite{kormanyos2015k} we set up a Hamiltonian in second quantization using layer localized eigenstates in the effective mass approximation as the basis. This Hamiltonian is then transformed into an exciton frame \cite{katsch2018theory,ivanov1993self,toyozawa1958theory} based on intra- and interlayer exciton states obtained from solving the bilayer Wannier equation \cite{ovesen2019interlayer,merkl2019ultrafast,laturia2018dielectric}. Finally, the excitonic Hamiltonian containing eigenenergies and interlayer interaction is diagonalized using a zone-folding approach \cite{ruiz2019interlayer} and focusing on the excitonic ground state. The interlayer interaction strength is extracted from the band splitting at different high-symmetry points in a perfectly aligned bilayer. The latter is calculated with density functional theory in the Perdew-Burke-Ernzerhof approximation, as implemented in the Quantum ESPRESSO package \cite{gillen2018interlayer}.  A detailed description of the developed microscopic approach to calculate energies and wave functions of layer hybridized intervalley moir\'e excitons in twisted bilayer TMDs is given in the supplementary material.

Figure \ref{fig:BS} shows the calculated exciton  bandstructure for K-K and K-$\Lambda$ excitons within their respective mBZ for tWSe$_2$ on a SiO$_2$ substrate. Here, the $\gamma$-point in the  K-$\Lambda$ mBZ corresponds to $\vec{Q}=\Lambda$. We have chosen two representative configurations: Figure \ref{fig:BS}(a) and (b) show K-K and K-$\Lambda$ dispersion at the 2$^{\circ}$ twist angle with the 3R stacking as reference (parallel, P stacking), while (c) and (d) correspond to the equivalent twist angle, however starting from the 2H configuration (anti-parallel, AP-stacking). The line colour reflects the projection of the exciton state onto an intralayer exciton. Here, red/blue corresponds to a pure intralayer/interlayer exciton, while green represents a 50/50 mix of both exciton types.
\begin{figure}[t!]
\includegraphics[width=\columnwidth]{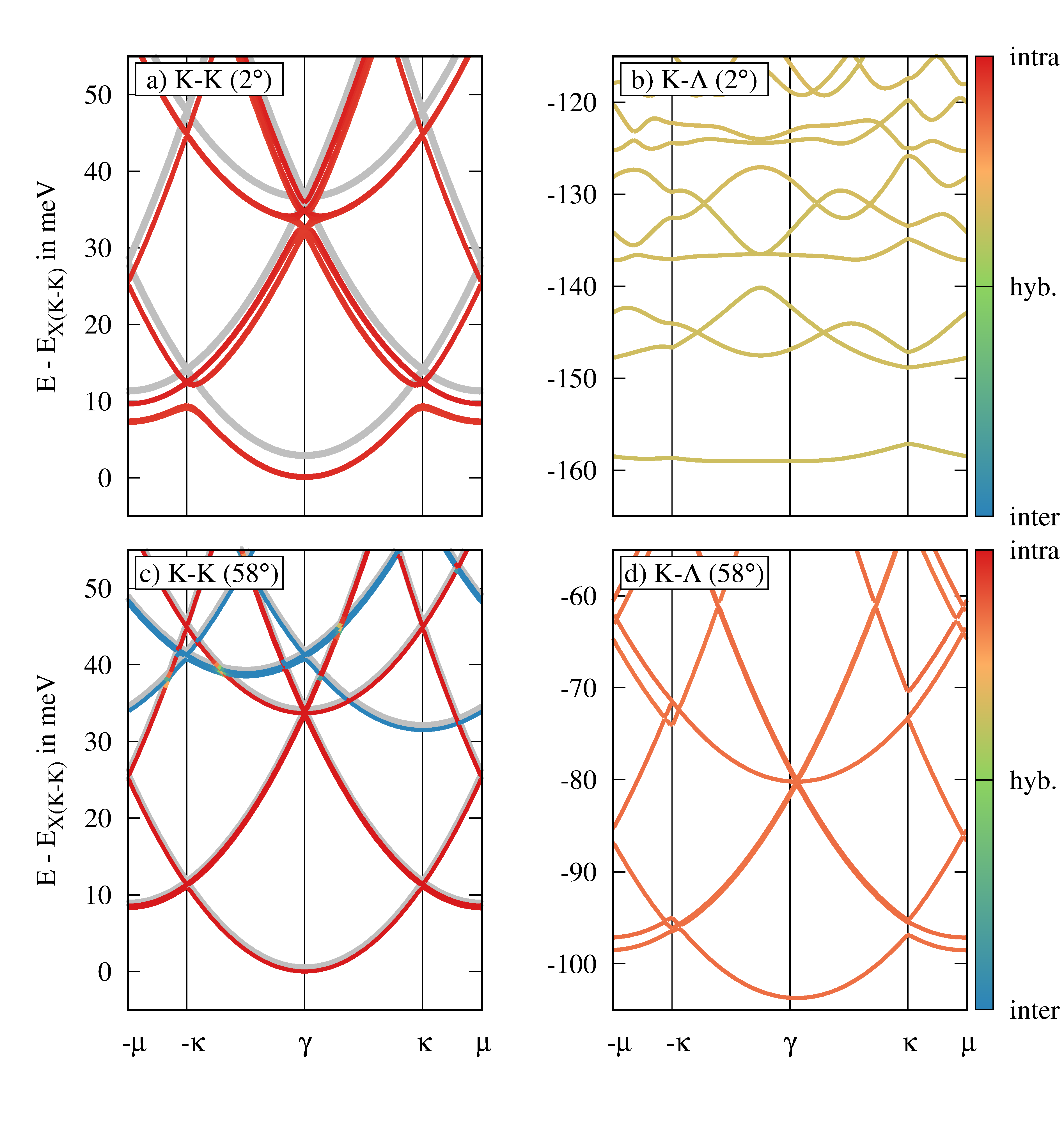}
\caption{Excitonic center-of-mass dispersion within the corresponding mBZ for 2$^{\circ}$ twisted (a)-(b) and 58$^{\circ}$ twisted (c)-(d) bilayer WSe$_2$ on SiO$_2$. For the bright K-K excitons [a,c] electrons and holes are tightly pinned to one of the two layers yielding either pure intra-(red) or interlayer exciton states (blue). The grey lines show the dispersion without taking into account hybridization. For the momentum-dark K-$\Lambda$  excitons [b,d] the pronounced delocalization of electrons at $\Lambda$ across both layers gives rise to a much stronger exciton hybridization (orange-green). Energies are plotted relative to the hybrid K-K exciton, respectively.}
\label{fig:BS} 
\end{figure}

First, we discuss the properties of K-K excitons [Fig.\ref{fig:BS}(a) and (c)]. For both stackings, we find very weak hybridization effects and the obtained dispersions are well described by the intra- and interlayer exciton energies without accounting for interlayer hybridization (grey lines). The microscopic reason is the small wave function overlap  of the adjacent layers. The electronic wave functions at the K point are mostly composed of d orbitals localized at tungsten atoms, which are sandwiched by the selenium atoms [Fig.\ref{fig:concept}(b)]. This results in a valence band tunnelling strength of about 10meV and a negligible coupling of the conduction bands (0.2meV).
In the case of P-stacking (2$^{\circ}$), the electronic bands of both layers are initially degenerate [Fig. \ref{fig:concept}(c)]. However, when accounting for the Coulomb interaction, we find that the interlayer exciton has a much weaker binding energy than the intralayer exciton, so that the minima of their respective parabolas are energetically separated by about 75meV  [Fig. \ref{fig:concept}(d)]. This detuning suppresses hybridization and the resulting redshift of the ground state is less than 5meV.
For AP-stacking (58$^{\circ}$), the ordering of the spin-split bands in the two layers is inverted. Therefore, the hole-hopping is entirely blocked due to the large spin-orbit-coupling in the valence band, resulting in a negligible red-shift. 

Now, we discuss the dispersion of K-$\Lambda$ excitons [Fig.\ref{fig:BS}(b) and (d)]. Here, we find a very strong hybridization of K-$\Lambda$ intra- and interlayer excitons in particular for P-stacking (2$^{\circ}$). The conduction band wave function at $\Lambda$ has large contributions at the selenium atoms [Fig.\ref{fig:concept}(b)], so that the hopping integral is one order of magnitude larger (T$\sim$170meV) than that of the K point, giving rise to an efficient mixing.
For P-stacking, the delocalization of electrons at the $\Lambda$ valley leads to a redshift of about 125meV (non-hybridized K-$\Lambda$ states not shown), while the splitting of different moir\'e subbands in the range of 10meV is entirely mediated by the holes at the K point. Considering both electron and hole hopping, combined with the large effective mass at the $\Lambda$ point, this gives rise to the emergence of an almost flat K-$\Lambda$ band at approximately 160meV below the bright K-K exciton. The observed opening of an excitonic gap is a result of the periodic potential influencing the exciton center-of-mass motion. Considering the long lifetime of the optically dark K-$\Lambda$ excitons, the flat bandstructure can potentially facilitate excitonic Bose-Einstein condensation, exciton super fluidity and other exotic bosonic states \cite{zheng2014exotic,wu2018hubbard}. Moreover, flat bands correspond to vanishing exciton group velocities and can therefore also be interpreted as the emergence of moir\'e trapped exciton states. 

Finally, when considering K-$\Lambda$ excitons for AP-stacking, we find that both the hybridization, as well as the splitting of moir\'e subbands becomes strongly reduced.  The large spin-orbit-coupling in the conduction band at the $\Lambda$ point and in the valence band at the K point strongly quenches the interlayer hopping. However, despite the large separation between intra- and interlayer exciton, the strong overlap of electronic wave functions at the  $\Lambda$ valley still leads to a significant delocalization of electrons resulting in a red-shift of the K-$\Lambda$ exciton by about 100meV.  
  
Several theoretical as well as experimental studies have previously suggested that already in the case of WSe$_2$ monolayers, the large mass of $\Lambda$ electrons gives rise to an enhanced exciton binding energy, so that the K-$\Lambda$ exciton can be energetically below the bright K-K exciton \cite{selig2016excitonic, zhang2015experimental,niehues2018strain,ye2016pressure, brem2020phonon}. However, due to the comparably small energy difference between K-K and K-$\Lambda$ and the existence other lower lying dark states, the presence of the K-$\Lambda$ exciton is often neglected in literature. In agreement with recent ab initio studies \cite{deilmann2019finite}, we find that in the case of tWSe$_2$ the strong interlayer hybridization of  $\Lambda$ electrons leads to a drastic redshift of the K-$\Lambda$ exciton, so that it is by far the lowest lying exciton state. Furthermore, we find that, in contrast to the bright K-K excitons, hybridization and moir\'e effects dominate the properties of the momentum-dark K-$\Lambda$ excitons in tWSe$_2$.

\textbf{Absorption spectra.}
Now, we investigate how the hybridization and moir\'e effects discussed above impact the optical fingerprint of tWSe$_2$. We first perform microscopic calculations of absorption spectra. We derive a Hamilton operator for the exciton-light interaction of layer-hybridized moir\'e excitons, cf. the supplementary material.  Based on the new matrix elements, we  generalize the excitonic Elliot formula \cite{kira2011semiconductor}, so that the absorption coefficient $\alpha_\sigma$ for $\sigma$-polarized light reads
 \begin{eqnarray}\label{eq:elliot}
\alpha_\sigma(E)=\sum_{\nu}\Im\text{m}\bigg[\dfrac{A_{\nu\sigma}}{\mathcal{E}_{\nu\vec{0}}-E-\mi(\gamma^{\zeta}_{\nu} +\Gamma^{\zeta}_{\nu})}\bigg],
\end{eqnarray}
where the sum is performed over all moir\'e subbands $\nu$, each contributing a Lorentzian response at its energy $\mathcal{E}_{\nu\vec{0}}$ at $\vec{Q}=0$ (the $\gamma$ point of the mBZ) characterized by the oscillator strength $A_{\nu\sigma}$ and a linewidth determined by the radiative and non-radiative dephasing ($\gamma^{\zeta}_{\nu}$ and $\Gamma^{\zeta}_{\nu}$). 
In a monolayer, only the exciton with zero center-of-mass momentum can interact with light, giving rise to a single optical resonance. In contrast, we find here that in principle the whole series of moir\'e subbands contributes to the optical response \cite{wu2017topological,jin2019observation}. 
\begin{figure}[t!]
\includegraphics[width=\columnwidth]{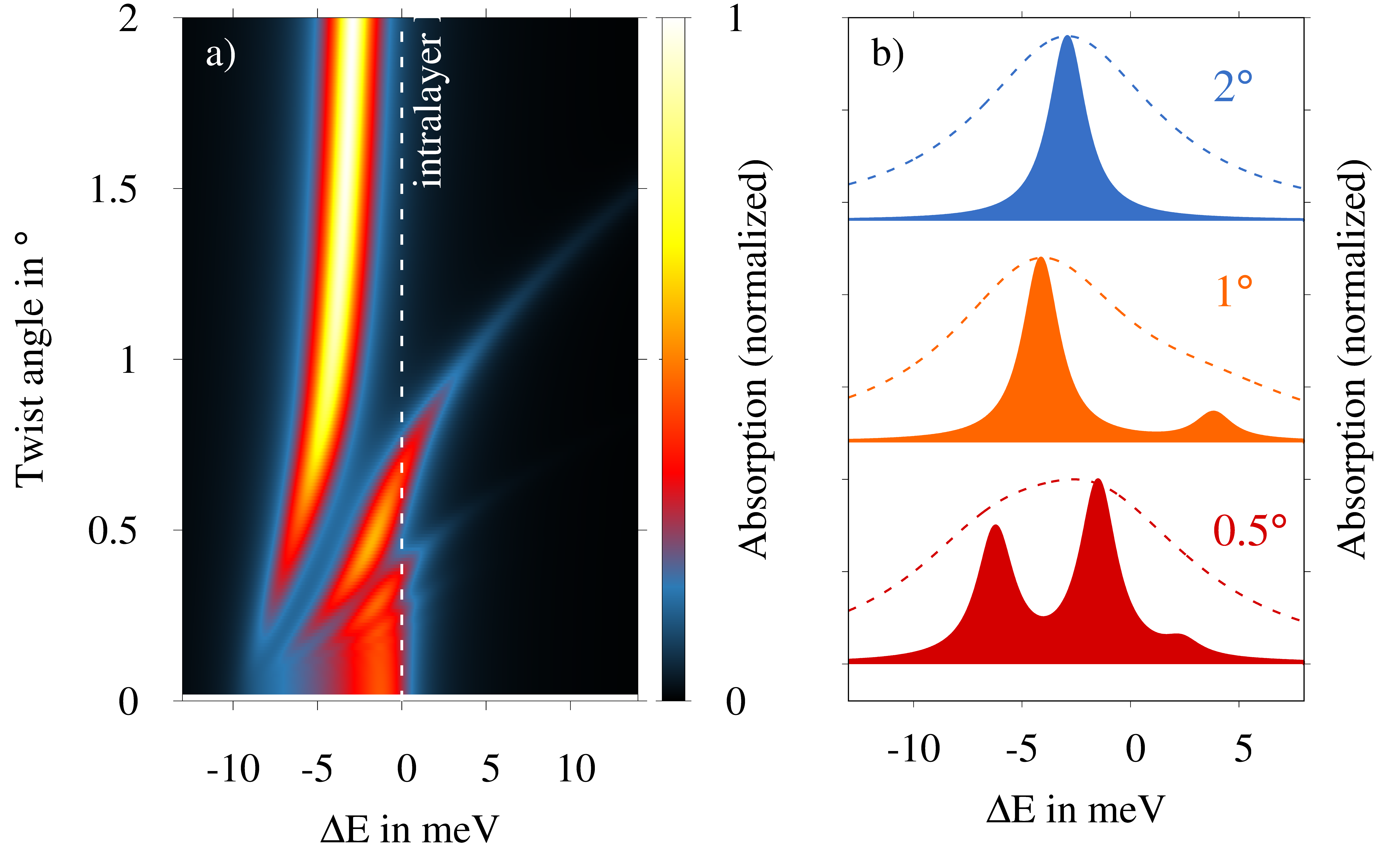}
\caption{Calculated twist-angle dependent absorption spectrum of a WSe$_2$ bilayer on SiO$_2$. Energies are plotted with respect to the intralayer exciton without considering interlayer hybridization [white dashed line in (a)]. While panel (a) shows the continuous evolution with twist angle, (b) contains cuts at different angles. The multiple resonances correspond to the moir\'e subbands at the $\gamma$ point in the mBZ (Fig. \ref{fig:concept}(e)), whose oscillator strength is given by their projection onto the bright $\vec{Q}=0$ excitons. The dashed lines in (b) illustrate spectra with a more realistic experimental excitonic linewidth of 5meV. }
\label{fig:absorp} 
\end{figure}
Figure \ref{fig:absorp} shows the twist-angle-dependent absorption spectrum in tWSe$_2$ for P-stacking, where hybridization and moir\'e effects are most pronounced.
For twist angles $\geq$2$^{\circ}$, we find a single resonance corresponding to the intralayer exciton (white dashed line), while its position is slightly red-shifted due to the weak interlayer hopping of K holes. When decreasing the twist angle to 1$^{\circ}$, the redshift increases and an additional peak appears. The position of this second peak is also moving down in energy with a quadratic dependence on the twist angle. However, instead of merging, the two peaks undergo an avoided crossing behaviour, while the oscillator strength is transferred to the higher-energy peak with further decreasing angles, cf. Fig. \ref{fig:absorp}(b). At even smaller twist angles $\leq$0.5$^{\circ}$, further high-energy moir\'e peaks appear including twist-angle-dependent red-shifts, avoided crossings and a transfer of oscillator strength. The appearance of additional peaks occurs in decreasing angle intervals so that the different resonances merge into a single peak at angles close to zero.

The microscopic origin of these additional moir\'e resonances is the modified momentum conservation in the moir\'e super lattice. In conventional semiconductors, only excitons with approximately zero center-of-mass momentum interact with light as the in-plane momentum provided by the photon is negligibly small. In contrast,  excitons in a superlattice can scatter with a moir\'e lattice vector \cite{ruiz2019interlayer} $\vec{b}_n=(C_3^n-1)\kappa$ (cf. Fig. \ref{fig:concept}a). Therefore, not only excitons with $\vec{Q}=0$ couple to light, but in principle all excitons with momenta coinciding with a $\gamma$ point of a certain mBZ do.
As a result, all exciton subbands in the centre of the mBZ obtain an oscillator strength, that however depends on how strongly they are hybridized with the original intralayer $\vec{Q}=0$ exciton.
We find that this oscillator strength is not pinned to the lowest lying exciton subband, but moves upwards in the series of subbands at $\gamma$ as the twist angle decreases. This finding is consistent with the fact that for $\theta\rightarrow0$, the mBZ collapses into a single point so that the number of states at $\gamma$ increases until they become continuous. At $\theta=0$, the oscillator strength of the continuous $\gamma$ states is distributed in such a way that the resulting optical response resembles the simple Lorentzian peak obtained for a perfectly aligned bilayer.

The predicted peak splittings are energetically small due to the weak interlayer coupling at the K point of WSe$_2$ and can only be resolved for very small exciton linewidths in the range of  $\Gamma=1$meV. In experiments performed at low temperatures and using hBN encapsulation, we still expect a larger dephasing due to efficient phonon emission and scattering into the lower lying K-$\Lambda$ states, so that the calculated moir\'e features will be smeared out into a single peak, cf. dashed lines in Fig. \ref{fig:absorp}(b) with linewidth $\Gamma=5$meV.

\textbf{Photoluminescence spectra.}
While linear optical transmission/absorption is only sensitive to the  bright (momentum-direct, spin-like, s-type) excitons, photoluminescence (PL) measurements in particular at low temperatures provide access to exciton states with very low oscillator strength. Here, the large occupation of the energetically lowest state as well as the long integration time of the photodetector enable us to measure signals stemming from more improbable transitions, such as the indirect phonon-assisted recombination of momentum-dark exciton states.
In a recent work \cite{brem2020phonon}, we developed a microscopic formalism for the phonon-assisted PL stemming from dark intervalley excitons. Here, we generalize this approach to interlayer hybrid moir\'e excitons (cf. the supplementary material) and show how the emission spectrum of tWSe$_2$ bilayer evolves with the twist angle. For the photon emission signal $I_\sigma$ perpendicular to the bilayer we find
\begin{eqnarray} \label{eq:phonon-PL} 
 I_\sigma(E)\propto\sum_{\zeta\nu} \hat{A}^{\zeta}_{\nu\sigma}(E) \bigg{(}  \gamma^{\zeta}_{\nu} N^\zeta_{\nu\vec{0}} + \sum_{\zeta'\nu'\mathbf{q}} \hat{B}^{\zeta'\zeta}_{\nu'\nu\vec{q}}(E)N^{\zeta'}_{\nu'\vec{q}}\bigg{)}.
\end{eqnarray}
with $\hat{A}^{\zeta}_{\nu\sigma}(E)= A^{\zeta}_{\nu\sigma}[(\mathcal{E}^{\zeta}_{\nu\vec{0}}-E)^2+(\gamma^{\zeta}_{\nu} +\Gamma^{\zeta}_{\nu})^2]^{-1}$ and $\hat{B}^{\zeta'\zeta}_{\nu'\nu\vec{q}}(E)= \Im\text{m}\sum_{\alpha\pm} |D^{\zeta'\zeta}_{\nu'\nu}(\alpha,\vec{q})|^2  \eta^{\pm}_{\alpha\mathbf{q}} [\mathcal{E}^{\zeta'}_{\nu'\vec{q}}\pm\hbar\Omega_{\alpha\mathbf{q}} -E - i\Gamma^{\zeta'}_{\nu'}]^{-1}$.
Here, we have introduced the exciton valley index $\zeta=K-K,K-\Lambda,K-K',...$ and the exciton occupation density $N^\zeta_{\nu\vec{Q}}$. Moreover, $\Omega_{\alpha\mathbf{q}}$ denotes the phonon frequency of mode $\alpha$ at momentum $\vec{q}$ and the phonon occupation factor $\eta^\pm_{\alpha\vec{q}}=1/2\mp1/2+n_B(\Omega_\vec{q})$ involves the Bose distribution $n_B$, weighting phonon absorption (+) and emission (-) processes.  
While the first part in Eq. \ref{eq:phonon-PL} contains direct radiative decay, the second part accounts for phonon-assisted recombination. Here, an exciton in state $(\zeta',\nu')$ scatters to a virtual state within the light cone ($\vec{Q} \approx 0$) via emission or absorption of a phonon and subsequently emits a photon with the remaining energy. The exciton-phonon scattering strength is determined by the effective matrix $D$ \cite{jin2014intrinsic,selig2018dark,brem2018exciton}, which is integrated over all possible moir\'e Umklapp processes \cite{lin2018moire}, cf. the supplementary material.

\begin{figure}[t!]
\includegraphics[width=\columnwidth]{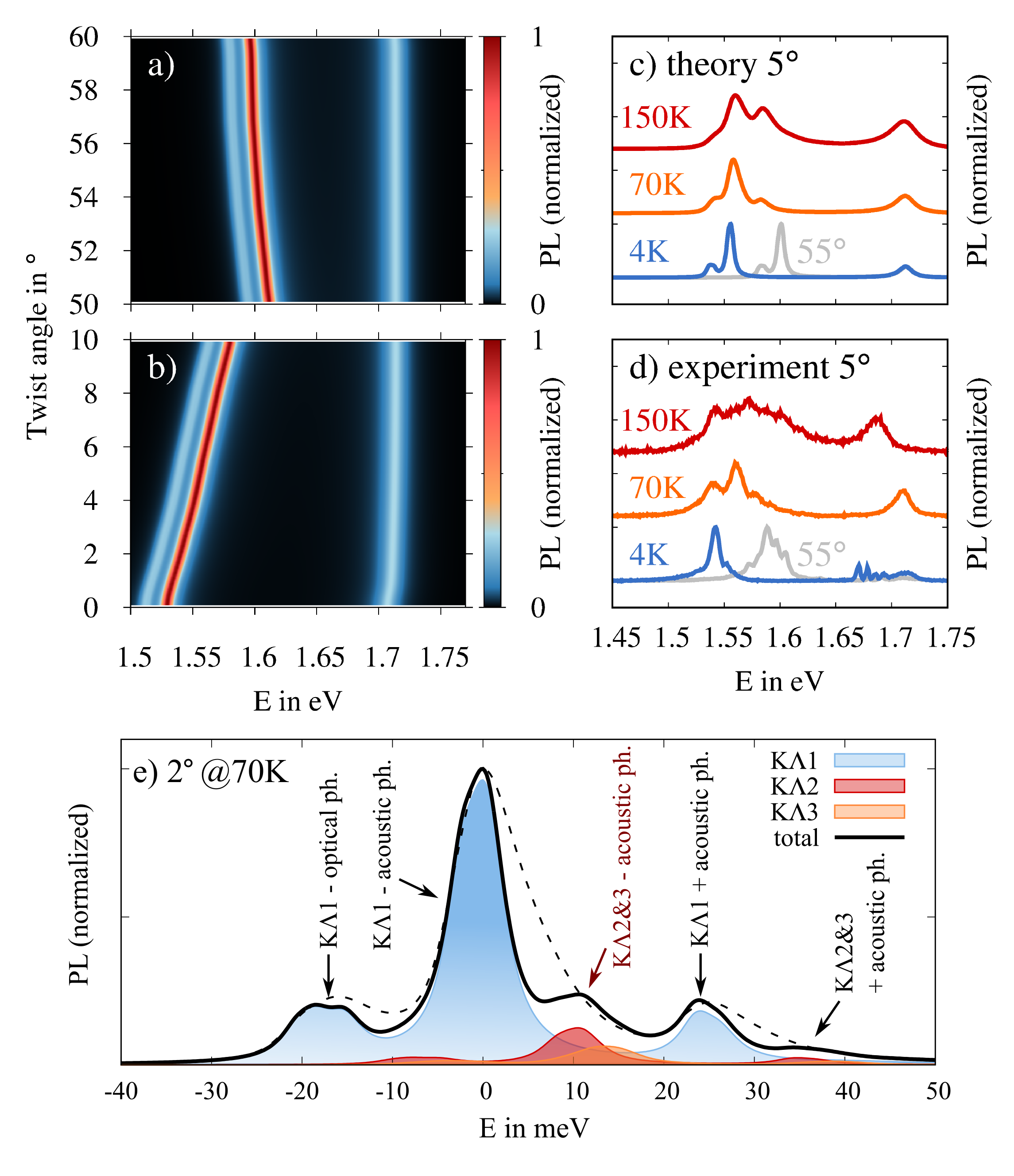}
\caption{Photoluminescence (PL) spectra of a twisted WSe$_2$ on SiO$_2$.  Twist-angle dependent PL calculated for (a) anti-parallel and  (b) parallel stacking at 4K. Direct comparison of (c) calculated and (d)  experimentally measured PL for different temperatures at a fixed angle of 5$^{\circ}$. The gray curves show the spectra at 4K and 55$^{\circ}$. (e) Phonon-assisted PL of K-$\Lambda$ excitons at 2$^{\circ}$ and 70K. We predict a symmetric phonon-sideband and high energy shoulders resulting from the flat exciton bands and higher-order moir\'e subbands (Fig. \ref{fig:BS}(b)). The dashed curve shows 5$^{\circ}$  PL for comparison.}
\label{fig:PL} 
\end{figure}

Figure \ref{fig:PL} shows the twist-angle and temperature-dependent PL of tWSe$_2$. Both for P- and AP-stacking, we find two distinct PL signals. The energetically higher peak at around 1.7 eV is independent  of the twist angle and reflects the recombination of the weakly hybridized bright K-K excitons.  In contrast, the energetically lower peak undergoes a twist angle-dependent shift in the range of several 10s of meV. This peak is a result of phonon-assisted photoluminescence of the momentum-dark K-$\Lambda$ excitons. Since the energy of the phonons involved changes only weakly with the stacking angle \cite{lin2018moire}, the evolution of the peak directly reflects the energetic shift of the K-$\Lambda$ exciton induced by interlayer hybridization.
There are two microscopic aspects, which give rise to a decreased hybridization resulting in the predicted blue-shift when twisting the bilayer away from the commensurate alignment at 0 and 60$^{\circ}$. Increasing the twist angle leads to an increased momentum shift of intra- and interlayer exciton dispersion. This results in (i) an increased detuning between intra- and interlayer dispersions at the $\gamma$ point in the mBZ, and (ii) a reduced excitonic hopping integral due to a decreased in-plane overlap of excitonic wave functions.   
Moreover, we find a clear asymmetry between the P and AP stacking. This asymmetry is a result of the inverted ordering of spin-split states in the two adjacent layers for the AP-configuration resulting in a suppressed hybridization as discussed in Fig. \ref{fig:BS}.

In Figs. \ref{fig:PL}(a) and (b) the temperature is very low (4K) and therefore we only observe two K-$\Lambda$ resonances, stemming from recombination of excitons in the lowest moir\'e subband assisted by the emission of either optical or acoustic $\Lambda$ phonons. Figure \ref{fig:PL}(c) shows the calculated PL spectrum at a fixed twist angle of 5$^{\circ}$ at three different temperatures. With increasing temperature all signals become broader due to enhanced phonon-induced dephasing, which was phenomenologically chosen to increase linearly with temperature (cf. supplementary material). Moreover, the exciton distribution becomes broader in energy, giving rise to an enhanced high energy tail of the K-$\Lambda$ peak and a relative increase of the direct K-K emission. Note that the K-K exciton peak remains symmetric at all temperatures as a result of the momentum selection rules. These are absent for K-$\Lambda$ excitons due to the assistance of a phonon, so that the shape of the phonon side band reflects the energetic distribution of K-$\Lambda$ excitons. Finally, at higher temperatures we predict the emergence of an additional peak at approximately 1.58eV. This peak is related to the decay of a K-$\Lambda$ exciton assisted by the absorption of an acoustic phonon, the probability of which is proportional to the phonon occupation number.

Now, we directly compare the theoretically predicted with  experimentally measured PL of a WSe$_2$ bilayer twisted by 5$^{\circ}$, cf. Fig.\ref{fig:PL}(d).
Details on the sample preparation and experimental set-up can be found in the supplementary. We find a good agreement between theory and experiment in terms of the energetic positions of excitonic resonances as well as the evolution with temperature of the line shapes. In particular,  the emergence of the asymmetric broadening towards higher energies  is a strong indicator for the correct microscopic assignment of the low energy peak to a phonon-assisted recombination of K-$\Lambda$ excitons. Moreover, when comparing 5 and 55$^{\circ}$ (grey curve) measurements we also find a similar shift of the K-$\Lambda$ peak in theory and experiment, which further verifies the validity of our microscopic model. 

Finally, we consider the case of low twist angles and intermediate temperatures allowing us to capture moir\'e features of dark intervalley excitons. Fig.\ref{fig:PL}(e) displays the calculated PL spectrum at 2$^{\circ}$ and 70K (black solid line) in the vicinity of the largest phonon side peak. Just as in the case of 5$^{\circ}$ we find three phonon-assisted peaks reflecting the decay of the energetically lowest K-$\Lambda$ moir\'e band (K$\Lambda$1), assisted by the emission of acoustic (-20meV) and optical $\Lambda$ phonons (main peak at 0meV)  or the absorption of an acoustic phonon (+25meV). However, when comparing the emission signal with the 5$^{\circ}$ spectrum (dashed line) we find two important differences: (i) while the 5$^{\circ}$ peaks are asymmetrically shaped reflecting the Boltzmann exciton distribution, the peaks at 2$^{\circ}$ are much narrower and appear fully symmetric. This reflects the almost discrete density of states for the flat K-$\Lambda$ moir\'e band predicted at 2$^{\circ}$, cf. Fig. \ref{fig:BS}(b). (ii) We predict the appearance of an additional peak at about 10meV above the main peak, which results from the occupation of higher order moir\'e bands (K$\Lambda$2 and K$\Lambda$3) decaying through emission of an acoustic phonon. The contributions of  different moir\'e bands to the overall PL signal are shown as coloured curves. This additional peak is an indicator for the quantization of the center-of-mass motion, since it reflects a discontinuity in the density of states.  Additionally, the symmetric line shape of the phonon side band can be used as an indicator for the existence of moir\'e trapped dark intervalley excitons.  

\textbf{Conclusion.} 
In summary, we have developed a microscopic model combining density-matrix and density-functional theory to calculate the excitonic bandstructure as well as absorption and PL spectra in twisted TMD homobilayers. Our approach explicitly includes the emergence of momentum-direct and indirect intra- and interlayer excitons and accounts for hybridization and moir\'e effects. We find that in contrast to the excitons at the K point, the energetically lowest K-$\Lambda$ excitons are strongly delocalized over both layers resulting in pronounced hybridization effects. The phonon-sidebands of these momentum-dark excitons in the low temperature PL provide direct access to the twist-angle dependent interlayer hybridization. Moreover, for small twist angles we predict a symmetric phonon-sideband and high energy shoulders resulting from the flat exciton bands and higher order moir\'e subbands, respectively. 
Our work contributes to a better understanding of exciton and moir\'e physics in van-der-Waals stacked multilayer systems and will trigger future studies on hybrid excitons in twisted 2D materials. 
\\

\textbf{Conflicts of interest}
There are no conflicts of interest to declare.

\begin{acknowledgments}
The Chalmers group acknowledges financial support from the European Unions Horizon 2020 research and innovation program under grant agreement No 785219 (Graphene Flagship) as well as from the Swedish Research Council (VR, project number 2018-00734). The Regensburg group acknowledges financial support from Deutsche Forschungsgemeinschaft (DFG, German Research Foundation) - Project-ID 314695032 - SFB 1277 project B03.
\end{acknowledgments}


\begin{thebibliography}{50}%
\makeatletter
\providecommand \@ifxundefined [1]{%
 \@ifx{#1\undefined}
}%
\providecommand \@ifnum [1]{%
 \ifnum #1\expandafter \@firstoftwo
 \else \expandafter \@secondoftwo
 \fi
}%
\providecommand \@ifx [1]{%
 \ifx #1\expandafter \@firstoftwo
 \else \expandafter \@secondoftwo
 \fi
}%
\providecommand \natexlab [1]{#1}%
\providecommand \enquote  [1]{``#1''}%
\providecommand \bibnamefont  [1]{#1}%
\providecommand \bibfnamefont [1]{#1}%
\providecommand \citenamefont [1]{#1}%
\providecommand \href@noop [0]{\@secondoftwo}%
\providecommand \href [0]{\begingroup \@sanitize@url \@href}%
\providecommand \@href[1]{\@@startlink{#1}\@@href}%
\providecommand \@@href[1]{\endgroup#1\@@endlink}%
\providecommand \@sanitize@url [0]{\catcode `\\12\catcode `\$12\catcode
  `\&12\catcode `\#12\catcode `\^12\catcode `\_12\catcode `\%12\relax}%
\providecommand \@@startlink[1]{}%
\providecommand \@@endlink[0]{}%
\providecommand \url  [0]{\begingroup\@sanitize@url \@url }%
\providecommand \@url [1]{\endgroup\@href {#1}{\urlprefix }}%
\providecommand \urlprefix  [0]{URL }%
\providecommand \Eprint [0]{\href }%
\providecommand \doibase [0]{http://dx.doi.org/}%
\providecommand \selectlanguage [0]{\@gobble}%
\providecommand \bibinfo  [0]{\@secondoftwo}%
\providecommand \bibfield  [0]{\@secondoftwo}%
\providecommand \translation [1]{[#1]}%
\providecommand \BibitemOpen [0]{}%
\providecommand \bibitemStop [0]{}%
\providecommand \bibitemNoStop [0]{.\EOS\space}%
\providecommand \EOS [0]{\spacefactor3000\relax}%
\providecommand \BibitemShut  [1]{\csname bibitem#1\endcsname}%
\let\auto@bib@innerbib\@empty
\bibitem [{\citenamefont {Geim}\ and\ \citenamefont
  {Grigorieva}(2013)}]{geim2013van}%
  \BibitemOpen
  \bibfield  {author} {\bibinfo {author} {\bibfnamefont {A.~K.}\ \bibnamefont
  {Geim}}\ and\ \bibinfo {author} {\bibfnamefont {I.~V.}\ \bibnamefont
  {Grigorieva}},\ }\href@noop {} {\bibfield  {journal} {\bibinfo  {journal}
  {Nature}\ }\textbf {\bibinfo {volume} {499}},\ \bibinfo {pages} {419}
  (\bibinfo {year} {2013})}\BibitemShut {NoStop}%
\bibitem [{\citenamefont {Kunstmann}\ \emph {et~al.}(2018)\citenamefont
  {Kunstmann}, \citenamefont {Mooshammer}, \citenamefont {Nagler},
  \citenamefont {Chaves}, \citenamefont {Stein}, \citenamefont {Paradiso},
  \citenamefont {Plechinger}, \citenamefont {Strunk}, \citenamefont
  {Sch{\"u}ller}, \citenamefont {Seifert} \emph
  {et~al.}}]{kunstmann2018momentum}%
  \BibitemOpen
  \bibfield  {author} {\bibinfo {author} {\bibfnamefont {J.}~\bibnamefont
  {Kunstmann}}, \bibinfo {author} {\bibfnamefont {F.}~\bibnamefont
  {Mooshammer}}, \bibinfo {author} {\bibfnamefont {P.}~\bibnamefont {Nagler}},
  \bibinfo {author} {\bibfnamefont {A.}~\bibnamefont {Chaves}}, \bibinfo
  {author} {\bibfnamefont {F.}~\bibnamefont {Stein}}, \bibinfo {author}
  {\bibfnamefont {N.}~\bibnamefont {Paradiso}}, \bibinfo {author}
  {\bibfnamefont {G.}~\bibnamefont {Plechinger}}, \bibinfo {author}
  {\bibfnamefont {C.}~\bibnamefont {Strunk}}, \bibinfo {author} {\bibfnamefont
  {C.}~\bibnamefont {Sch{\"u}ller}}, \bibinfo {author} {\bibfnamefont
  {G.}~\bibnamefont {Seifert}},  \emph {et~al.},\ }\href@noop {} {\bibfield
  {journal} {\bibinfo  {journal} {Nature Physics}\ }\textbf {\bibinfo {volume}
  {14}},\ \bibinfo {pages} {801} (\bibinfo {year} {2018})}\BibitemShut
  {NoStop}%
\bibitem [{\citenamefont {Zhang}\ \emph {et~al.}(2018)\citenamefont {Zhang},
  \citenamefont {Surrente}, \citenamefont {Baranowski}, \citenamefont {Maude},
  \citenamefont {Gant}, \citenamefont {Castellanos-Gomez},\ and\ \citenamefont
  {Plochocka}}]{zhang2018moire}%
  \BibitemOpen
  \bibfield  {author} {\bibinfo {author} {\bibfnamefont {N.}~\bibnamefont
  {Zhang}}, \bibinfo {author} {\bibfnamefont {A.}~\bibnamefont {Surrente}},
  \bibinfo {author} {\bibfnamefont {M.}~\bibnamefont {Baranowski}}, \bibinfo
  {author} {\bibfnamefont {D.~K.}\ \bibnamefont {Maude}}, \bibinfo {author}
  {\bibfnamefont {P.}~\bibnamefont {Gant}}, \bibinfo {author} {\bibfnamefont
  {A.}~\bibnamefont {Castellanos-Gomez}}, \ and\ \bibinfo {author}
  {\bibfnamefont {P.}~\bibnamefont {Plochocka}},\ }\href@noop {} {\bibfield
  {journal} {\bibinfo  {journal} {Nano letters}\ }\textbf {\bibinfo {volume}
  {18}},\ \bibinfo {pages} {7651} (\bibinfo {year} {2018})}\BibitemShut
  {NoStop}%
\bibitem [{\citenamefont {Merkl}\ \emph {et~al.}(2019)\citenamefont {Merkl},
  \citenamefont {Mooshammer}, \citenamefont {Steinleitner}, \citenamefont
  {Girnghuber}, \citenamefont {Lin}, \citenamefont {Nagler}, \citenamefont
  {Holler}, \citenamefont {Sch{\"u}ller}, \citenamefont {Lupton}, \citenamefont
  {Korn} \emph {et~al.}}]{merkl2019ultrafast}%
  \BibitemOpen
  \bibfield  {author} {\bibinfo {author} {\bibfnamefont {P.}~\bibnamefont
  {Merkl}}, \bibinfo {author} {\bibfnamefont {F.}~\bibnamefont {Mooshammer}},
  \bibinfo {author} {\bibfnamefont {P.}~\bibnamefont {Steinleitner}}, \bibinfo
  {author} {\bibfnamefont {A.}~\bibnamefont {Girnghuber}}, \bibinfo {author}
  {\bibfnamefont {K.-Q.}\ \bibnamefont {Lin}}, \bibinfo {author} {\bibfnamefont
  {P.}~\bibnamefont {Nagler}}, \bibinfo {author} {\bibfnamefont
  {J.}~\bibnamefont {Holler}}, \bibinfo {author} {\bibfnamefont
  {C.}~\bibnamefont {Sch{\"u}ller}}, \bibinfo {author} {\bibfnamefont {J.~M.}\
  \bibnamefont {Lupton}}, \bibinfo {author} {\bibfnamefont {T.}~\bibnamefont
  {Korn}},  \emph {et~al.},\ }\href@noop {} {\bibfield  {journal} {\bibinfo
  {journal} {Nature materials}\ }\textbf {\bibinfo {volume} {18}},\ \bibinfo
  {pages} {691} (\bibinfo {year} {2019})}\BibitemShut {NoStop}%
\bibitem [{\citenamefont {Dean}\ \emph {et~al.}(2010)\citenamefont {Dean},
  \citenamefont {Young}, \citenamefont {Meric}, \citenamefont {Lee},
  \citenamefont {Wang}, \citenamefont {Sorgenfrei}, \citenamefont {Watanabe},
  \citenamefont {Taniguchi}, \citenamefont {Kim}, \citenamefont {Shepard} \emph
  {et~al.}}]{dean2010boron}%
  \BibitemOpen
  \bibfield  {author} {\bibinfo {author} {\bibfnamefont {C.~R.}\ \bibnamefont
  {Dean}}, \bibinfo {author} {\bibfnamefont {A.~F.}\ \bibnamefont {Young}},
  \bibinfo {author} {\bibfnamefont {I.}~\bibnamefont {Meric}}, \bibinfo
  {author} {\bibfnamefont {C.}~\bibnamefont {Lee}}, \bibinfo {author}
  {\bibfnamefont {L.}~\bibnamefont {Wang}}, \bibinfo {author} {\bibfnamefont
  {S.}~\bibnamefont {Sorgenfrei}}, \bibinfo {author} {\bibfnamefont
  {K.}~\bibnamefont {Watanabe}}, \bibinfo {author} {\bibfnamefont
  {T.}~\bibnamefont {Taniguchi}}, \bibinfo {author} {\bibfnamefont
  {P.}~\bibnamefont {Kim}}, \bibinfo {author} {\bibfnamefont {K.~L.}\
  \bibnamefont {Shepard}},  \emph {et~al.},\ }\href@noop {} {\bibfield
  {journal} {\bibinfo  {journal} {Nature nanotechnology}\ }\textbf {\bibinfo
  {volume} {5}},\ \bibinfo {pages} {722} (\bibinfo {year} {2010})}\BibitemShut
  {NoStop}%
\bibitem [{\citenamefont {Raja}\ \emph {et~al.}(2019)\citenamefont {Raja},
  \citenamefont {Waldecker}, \citenamefont {Zipfel}, \citenamefont {Cho},
  \citenamefont {Brem}, \citenamefont {Ziegler}, \citenamefont {Kulig},
  \citenamefont {Taniguchi}, \citenamefont {Watanabe}, \citenamefont {Malic}
  \emph {et~al.}}]{raja2019dielectric}%
  \BibitemOpen
  \bibfield  {author} {\bibinfo {author} {\bibfnamefont {A.}~\bibnamefont
  {Raja}}, \bibinfo {author} {\bibfnamefont {L.}~\bibnamefont {Waldecker}},
  \bibinfo {author} {\bibfnamefont {J.}~\bibnamefont {Zipfel}}, \bibinfo
  {author} {\bibfnamefont {Y.}~\bibnamefont {Cho}}, \bibinfo {author}
  {\bibfnamefont {S.}~\bibnamefont {Brem}}, \bibinfo {author} {\bibfnamefont
  {J.~D.}\ \bibnamefont {Ziegler}}, \bibinfo {author} {\bibfnamefont
  {M.}~\bibnamefont {Kulig}}, \bibinfo {author} {\bibfnamefont
  {T.}~\bibnamefont {Taniguchi}}, \bibinfo {author} {\bibfnamefont
  {K.}~\bibnamefont {Watanabe}}, \bibinfo {author} {\bibfnamefont
  {E.}~\bibnamefont {Malic}},  \emph {et~al.},\ }\href@noop {} {\bibfield
  {journal} {\bibinfo  {journal} {Nature nanotechnology}\ }\textbf {\bibinfo
  {volume} {14}},\ \bibinfo {pages} {832} (\bibinfo {year} {2019})}\BibitemShut
  {NoStop}%
\bibitem [{\citenamefont {Fang}\ \emph {et~al.}(2014)\citenamefont {Fang},
  \citenamefont {Battaglia}, \citenamefont {Carraro}, \citenamefont {Nemsak},
  \citenamefont {Ozdol}, \citenamefont {Kang}, \citenamefont {Bechtel},
  \citenamefont {Desai}, \citenamefont {Kronast}, \citenamefont {Unal} \emph
  {et~al.}}]{fang2014strong}%
  \BibitemOpen
  \bibfield  {author} {\bibinfo {author} {\bibfnamefont {H.}~\bibnamefont
  {Fang}}, \bibinfo {author} {\bibfnamefont {C.}~\bibnamefont {Battaglia}},
  \bibinfo {author} {\bibfnamefont {C.}~\bibnamefont {Carraro}}, \bibinfo
  {author} {\bibfnamefont {S.}~\bibnamefont {Nemsak}}, \bibinfo {author}
  {\bibfnamefont {B.}~\bibnamefont {Ozdol}}, \bibinfo {author} {\bibfnamefont
  {J.~S.}\ \bibnamefont {Kang}}, \bibinfo {author} {\bibfnamefont {H.~A.}\
  \bibnamefont {Bechtel}}, \bibinfo {author} {\bibfnamefont {S.~B.}\
  \bibnamefont {Desai}}, \bibinfo {author} {\bibfnamefont {F.}~\bibnamefont
  {Kronast}}, \bibinfo {author} {\bibfnamefont {A.~A.}\ \bibnamefont {Unal}},
  \emph {et~al.},\ }\href@noop {} {\bibfield  {journal} {\bibinfo  {journal}
  {Proceedings of the National Academy of Sciences}\ }\textbf {\bibinfo
  {volume} {111}},\ \bibinfo {pages} {6198} (\bibinfo {year}
  {2014})}\BibitemShut {NoStop}%
\bibitem [{\citenamefont {Coy~Diaz}\ \emph {et~al.}(2015)\citenamefont
  {Coy~Diaz}, \citenamefont {Avila}, \citenamefont {Chen}, \citenamefont
  {Addou}, \citenamefont {Asensio},\ and\ \citenamefont
  {Batzill}}]{coy2015direct}%
  \BibitemOpen
  \bibfield  {author} {\bibinfo {author} {\bibfnamefont {H.}~\bibnamefont
  {Coy~Diaz}}, \bibinfo {author} {\bibfnamefont {J.}~\bibnamefont {Avila}},
  \bibinfo {author} {\bibfnamefont {C.}~\bibnamefont {Chen}}, \bibinfo {author}
  {\bibfnamefont {R.}~\bibnamefont {Addou}}, \bibinfo {author} {\bibfnamefont
  {M.~C.}\ \bibnamefont {Asensio}}, \ and\ \bibinfo {author} {\bibfnamefont
  {M.}~\bibnamefont {Batzill}},\ }\href@noop {} {\bibfield  {journal} {\bibinfo
   {journal} {Nano letters}\ }\textbf {\bibinfo {volume} {15}},\ \bibinfo
  {pages} {1135} (\bibinfo {year} {2015})}\BibitemShut {NoStop}%
\bibitem [{\citenamefont {Alexeev}\ \emph {et~al.}(2019)\citenamefont
  {Alexeev}, \citenamefont {Ruiz-Tijerina}, \citenamefont {Danovich},
  \citenamefont {Hamer}, \citenamefont {Terry}, \citenamefont {Nayak},
  \citenamefont {Ahn}, \citenamefont {Pak}, \citenamefont {Lee}, \citenamefont
  {Sohn} \emph {et~al.}}]{alexeev2019resonantly}%
  \BibitemOpen
  \bibfield  {author} {\bibinfo {author} {\bibfnamefont {E.~M.}\ \bibnamefont
  {Alexeev}}, \bibinfo {author} {\bibfnamefont {D.~A.}\ \bibnamefont
  {Ruiz-Tijerina}}, \bibinfo {author} {\bibfnamefont {M.}~\bibnamefont
  {Danovich}}, \bibinfo {author} {\bibfnamefont {M.~J.}\ \bibnamefont {Hamer}},
  \bibinfo {author} {\bibfnamefont {D.~J.}\ \bibnamefont {Terry}}, \bibinfo
  {author} {\bibfnamefont {P.~K.}\ \bibnamefont {Nayak}}, \bibinfo {author}
  {\bibfnamefont {S.}~\bibnamefont {Ahn}}, \bibinfo {author} {\bibfnamefont
  {S.}~\bibnamefont {Pak}}, \bibinfo {author} {\bibfnamefont {J.}~\bibnamefont
  {Lee}}, \bibinfo {author} {\bibfnamefont {J.~I.}\ \bibnamefont {Sohn}},
  \emph {et~al.},\ }\href@noop {} {\bibfield  {journal} {\bibinfo  {journal}
  {Nature}\ }\textbf {\bibinfo {volume} {567}},\ \bibinfo {pages} {81}
  (\bibinfo {year} {2019})}\BibitemShut {NoStop}%
\bibitem [{\citenamefont {van Der~Zande}\ \emph {et~al.}(2014)\citenamefont
  {van Der~Zande}, \citenamefont {Kunstmann}, \citenamefont {Chernikov},
  \citenamefont {Chenet}, \citenamefont {You}, \citenamefont {Zhang},
  \citenamefont {Huang}, \citenamefont {Berkelbach}, \citenamefont {Wang},
  \citenamefont {Zhang} \emph {et~al.}}]{van2014tailoring}%
  \BibitemOpen
  \bibfield  {author} {\bibinfo {author} {\bibfnamefont {A.~M.}\ \bibnamefont
  {van Der~Zande}}, \bibinfo {author} {\bibfnamefont {J.}~\bibnamefont
  {Kunstmann}}, \bibinfo {author} {\bibfnamefont {A.}~\bibnamefont
  {Chernikov}}, \bibinfo {author} {\bibfnamefont {D.~A.}\ \bibnamefont
  {Chenet}}, \bibinfo {author} {\bibfnamefont {Y.}~\bibnamefont {You}},
  \bibinfo {author} {\bibfnamefont {X.}~\bibnamefont {Zhang}}, \bibinfo
  {author} {\bibfnamefont {P.~Y.}\ \bibnamefont {Huang}}, \bibinfo {author}
  {\bibfnamefont {T.~C.}\ \bibnamefont {Berkelbach}}, \bibinfo {author}
  {\bibfnamefont {L.}~\bibnamefont {Wang}}, \bibinfo {author} {\bibfnamefont
  {F.}~\bibnamefont {Zhang}},  \emph {et~al.},\ }\href@noop {} {\bibfield
  {journal} {\bibinfo  {journal} {Nano letters}\ }\textbf {\bibinfo {volume}
  {14}},\ \bibinfo {pages} {3869} (\bibinfo {year} {2014})}\BibitemShut
  {NoStop}%
\bibitem [{\citenamefont {Yeh}\ \emph {et~al.}(2016)\citenamefont {Yeh},
  \citenamefont {Jin}, \citenamefont {Zaki}, \citenamefont {Kunstmann},
  \citenamefont {Chenet}, \citenamefont {Arefe}, \citenamefont {Sadowski},
  \citenamefont {Dadap}, \citenamefont {Sutter}, \citenamefont {Hone} \emph
  {et~al.}}]{yeh2016direct}%
  \BibitemOpen
  \bibfield  {author} {\bibinfo {author} {\bibfnamefont {P.-C.}\ \bibnamefont
  {Yeh}}, \bibinfo {author} {\bibfnamefont {W.}~\bibnamefont {Jin}}, \bibinfo
  {author} {\bibfnamefont {N.}~\bibnamefont {Zaki}}, \bibinfo {author}
  {\bibfnamefont {J.}~\bibnamefont {Kunstmann}}, \bibinfo {author}
  {\bibfnamefont {D.}~\bibnamefont {Chenet}}, \bibinfo {author} {\bibfnamefont
  {G.}~\bibnamefont {Arefe}}, \bibinfo {author} {\bibfnamefont {J.~T.}\
  \bibnamefont {Sadowski}}, \bibinfo {author} {\bibfnamefont {J.~I.}\
  \bibnamefont {Dadap}}, \bibinfo {author} {\bibfnamefont {P.}~\bibnamefont
  {Sutter}}, \bibinfo {author} {\bibfnamefont {J.}~\bibnamefont {Hone}},  \emph
  {et~al.},\ }\href@noop {} {\bibfield  {journal} {\bibinfo  {journal} {Nano
  letters}\ }\textbf {\bibinfo {volume} {16}},\ \bibinfo {pages} {953}
  (\bibinfo {year} {2016})}\BibitemShut {NoStop}%
\bibitem [{\citenamefont {Yu}\ \emph {et~al.}(2017)\citenamefont {Yu},
  \citenamefont {Liu}, \citenamefont {Tang}, \citenamefont {Xu},\ and\
  \citenamefont {Yao}}]{yu2017moire}%
  \BibitemOpen
  \bibfield  {author} {\bibinfo {author} {\bibfnamefont {H.}~\bibnamefont
  {Yu}}, \bibinfo {author} {\bibfnamefont {G.-B.}\ \bibnamefont {Liu}},
  \bibinfo {author} {\bibfnamefont {J.}~\bibnamefont {Tang}}, \bibinfo {author}
  {\bibfnamefont {X.}~\bibnamefont {Xu}}, \ and\ \bibinfo {author}
  {\bibfnamefont {W.}~\bibnamefont {Yao}},\ }\href@noop {} {\bibfield
  {journal} {\bibinfo  {journal} {Science advances}\ }\textbf {\bibinfo
  {volume} {3}},\ \bibinfo {pages} {e1701696} (\bibinfo {year}
  {2017})}\BibitemShut {NoStop}%
\bibitem [{\citenamefont {Wu}\ \emph {et~al.}(2018{\natexlab{a}})\citenamefont
  {Wu}, \citenamefont {Lovorn},\ and\ \citenamefont
  {MacDonald}}]{wu2018theory}%
  \BibitemOpen
  \bibfield  {author} {\bibinfo {author} {\bibfnamefont {F.}~\bibnamefont
  {Wu}}, \bibinfo {author} {\bibfnamefont {T.}~\bibnamefont {Lovorn}}, \ and\
  \bibinfo {author} {\bibfnamefont {A.}~\bibnamefont {MacDonald}},\ }\href@noop
  {} {\bibfield  {journal} {\bibinfo  {journal} {Physical Review B}\ }\textbf
  {\bibinfo {volume} {97}},\ \bibinfo {pages} {035306} (\bibinfo {year}
  {2018}{\natexlab{a}})}\BibitemShut {NoStop}%
\bibitem [{\citenamefont {Rivera}\ \emph {et~al.}(2018)\citenamefont {Rivera},
  \citenamefont {Yu}, \citenamefont {Seyler}, \citenamefont {Wilson},
  \citenamefont {Yao},\ and\ \citenamefont {Xu}}]{rivera2018interlayer}%
  \BibitemOpen
  \bibfield  {author} {\bibinfo {author} {\bibfnamefont {P.}~\bibnamefont
  {Rivera}}, \bibinfo {author} {\bibfnamefont {H.}~\bibnamefont {Yu}}, \bibinfo
  {author} {\bibfnamefont {K.~L.}\ \bibnamefont {Seyler}}, \bibinfo {author}
  {\bibfnamefont {N.~P.}\ \bibnamefont {Wilson}}, \bibinfo {author}
  {\bibfnamefont {W.}~\bibnamefont {Yao}}, \ and\ \bibinfo {author}
  {\bibfnamefont {X.}~\bibnamefont {Xu}},\ }\href@noop {} {\bibfield  {journal}
  {\bibinfo  {journal} {Nature nanotechnology}\ }\textbf {\bibinfo {volume}
  {13}},\ \bibinfo {pages} {1004} (\bibinfo {year} {2018})}\BibitemShut
  {NoStop}%
\bibitem [{\citenamefont {Tran}\ \emph {et~al.}(2019)\citenamefont {Tran},
  \citenamefont {Moody}, \citenamefont {Wu}, \citenamefont {Lu}, \citenamefont
  {Choi}, \citenamefont {Kim}, \citenamefont {Rai}, \citenamefont {Sanchez},
  \citenamefont {Quan}, \citenamefont {Singh} \emph
  {et~al.}}]{tran2019evidence}%
  \BibitemOpen
  \bibfield  {author} {\bibinfo {author} {\bibfnamefont {K.}~\bibnamefont
  {Tran}}, \bibinfo {author} {\bibfnamefont {G.}~\bibnamefont {Moody}},
  \bibinfo {author} {\bibfnamefont {F.}~\bibnamefont {Wu}}, \bibinfo {author}
  {\bibfnamefont {X.}~\bibnamefont {Lu}}, \bibinfo {author} {\bibfnamefont
  {J.}~\bibnamefont {Choi}}, \bibinfo {author} {\bibfnamefont {K.}~\bibnamefont
  {Kim}}, \bibinfo {author} {\bibfnamefont {A.}~\bibnamefont {Rai}}, \bibinfo
  {author} {\bibfnamefont {D.~A.}\ \bibnamefont {Sanchez}}, \bibinfo {author}
  {\bibfnamefont {J.}~\bibnamefont {Quan}}, \bibinfo {author} {\bibfnamefont
  {A.}~\bibnamefont {Singh}},  \emph {et~al.},\ }\href@noop {} {\bibfield
  {journal} {\bibinfo  {journal} {Nature}\ }\textbf {\bibinfo {volume} {567}},\
  \bibinfo {pages} {71} (\bibinfo {year} {2019})}\BibitemShut {NoStop}%
\bibitem [{\citenamefont {Seyler}\ \emph {et~al.}(2019)\citenamefont {Seyler},
  \citenamefont {Rivera}, \citenamefont {Yu}, \citenamefont {Wilson},
  \citenamefont {Ray}, \citenamefont {Mandrus}, \citenamefont {Yan},
  \citenamefont {Yao},\ and\ \citenamefont {Xu}}]{seyler2019signatures}%
  \BibitemOpen
  \bibfield  {author} {\bibinfo {author} {\bibfnamefont {K.~L.}\ \bibnamefont
  {Seyler}}, \bibinfo {author} {\bibfnamefont {P.}~\bibnamefont {Rivera}},
  \bibinfo {author} {\bibfnamefont {H.}~\bibnamefont {Yu}}, \bibinfo {author}
  {\bibfnamefont {N.~P.}\ \bibnamefont {Wilson}}, \bibinfo {author}
  {\bibfnamefont {E.~L.}\ \bibnamefont {Ray}}, \bibinfo {author} {\bibfnamefont
  {D.~G.}\ \bibnamefont {Mandrus}}, \bibinfo {author} {\bibfnamefont
  {J.}~\bibnamefont {Yan}}, \bibinfo {author} {\bibfnamefont {W.}~\bibnamefont
  {Yao}}, \ and\ \bibinfo {author} {\bibfnamefont {X.}~\bibnamefont {Xu}},\
  }\href@noop {} {\bibfield  {journal} {\bibinfo  {journal} {Nature}\ }\textbf
  {\bibinfo {volume} {567}},\ \bibinfo {pages} {66} (\bibinfo {year}
  {2019})}\BibitemShut {NoStop}%
\bibitem [{\citenamefont {Wu}\ \emph {et~al.}(2018{\natexlab{b}})\citenamefont
  {Wu}, \citenamefont {Lovorn}, \citenamefont {Tutuc},\ and\ \citenamefont
  {MacDonald}}]{wu2018hubbard}%
  \BibitemOpen
  \bibfield  {author} {\bibinfo {author} {\bibfnamefont {F.}~\bibnamefont
  {Wu}}, \bibinfo {author} {\bibfnamefont {T.}~\bibnamefont {Lovorn}}, \bibinfo
  {author} {\bibfnamefont {E.}~\bibnamefont {Tutuc}}, \ and\ \bibinfo {author}
  {\bibfnamefont {A.~H.}\ \bibnamefont {MacDonald}},\ }\href@noop {} {\bibfield
   {journal} {\bibinfo  {journal} {Physical review letters}\ }\textbf {\bibinfo
  {volume} {121}},\ \bibinfo {pages} {026402} (\bibinfo {year}
  {2018}{\natexlab{b}})}\BibitemShut {NoStop}%
\bibitem [{\citenamefont {Wang}\ \emph {et~al.}(2019)\citenamefont {Wang},
  \citenamefont {Shih}, \citenamefont {Ghiotto}, \citenamefont {Xian},
  \citenamefont {Rhodes}, \citenamefont {Tan}, \citenamefont {Claassen},
  \citenamefont {Kennes}, \citenamefont {Bai}, \citenamefont {Kim} \emph
  {et~al.}}]{wang2019magic}%
  \BibitemOpen
  \bibfield  {author} {\bibinfo {author} {\bibfnamefont {L.}~\bibnamefont
  {Wang}}, \bibinfo {author} {\bibfnamefont {E.-M.}\ \bibnamefont {Shih}},
  \bibinfo {author} {\bibfnamefont {A.}~\bibnamefont {Ghiotto}}, \bibinfo
  {author} {\bibfnamefont {L.}~\bibnamefont {Xian}}, \bibinfo {author}
  {\bibfnamefont {D.~A.}\ \bibnamefont {Rhodes}}, \bibinfo {author}
  {\bibfnamefont {C.}~\bibnamefont {Tan}}, \bibinfo {author} {\bibfnamefont
  {M.}~\bibnamefont {Claassen}}, \bibinfo {author} {\bibfnamefont {D.~M.}\
  \bibnamefont {Kennes}}, \bibinfo {author} {\bibfnamefont {Y.}~\bibnamefont
  {Bai}}, \bibinfo {author} {\bibfnamefont {B.}~\bibnamefont {Kim}},  \emph
  {et~al.},\ }\href@noop {} {\bibfield  {journal} {\bibinfo  {journal} {arXiv
  preprint arXiv:1910.12147}\ } (\bibinfo {year} {2019})}\BibitemShut {NoStop}%
\bibitem [{\citenamefont {An}\ \emph {et~al.}(2019)\citenamefont {An},
  \citenamefont {Cai}, \citenamefont {Huang}, \citenamefont {Wu}, \citenamefont
  {Lin}, \citenamefont {Ying}, \citenamefont {Ye}, \citenamefont {Feng},\ and\
  \citenamefont {Wang}}]{an2019interaction}%
  \BibitemOpen
  \bibfield  {author} {\bibinfo {author} {\bibfnamefont {L.}~\bibnamefont
  {An}}, \bibinfo {author} {\bibfnamefont {X.}~\bibnamefont {Cai}}, \bibinfo
  {author} {\bibfnamefont {M.}~\bibnamefont {Huang}}, \bibinfo {author}
  {\bibfnamefont {Z.}~\bibnamefont {Wu}}, \bibinfo {author} {\bibfnamefont
  {J.}~\bibnamefont {Lin}}, \bibinfo {author} {\bibfnamefont {Z.}~\bibnamefont
  {Ying}}, \bibinfo {author} {\bibfnamefont {Z.}~\bibnamefont {Ye}}, \bibinfo
  {author} {\bibfnamefont {X.}~\bibnamefont {Feng}}, \ and\ \bibinfo {author}
  {\bibfnamefont {N.}~\bibnamefont {Wang}},\ }\href@noop {} {\bibfield
  {journal} {\bibinfo  {journal} {arXiv preprint arXiv:1907.03966}\ } (\bibinfo
  {year} {2019})}\BibitemShut {NoStop}%
\bibitem [{\citenamefont {Zhang}\ \emph {et~al.}(2019)\citenamefont {Zhang},
  \citenamefont {Wang}, \citenamefont {Watanabe}, \citenamefont {Taniguchi},
  \citenamefont {Ueno}, \citenamefont {Tutuc},\ and\ \citenamefont
  {LeRoy}}]{zhang2019flat}%
  \BibitemOpen
  \bibfield  {author} {\bibinfo {author} {\bibfnamefont {Z.}~\bibnamefont
  {Zhang}}, \bibinfo {author} {\bibfnamefont {Y.}~\bibnamefont {Wang}},
  \bibinfo {author} {\bibfnamefont {K.}~\bibnamefont {Watanabe}}, \bibinfo
  {author} {\bibfnamefont {T.}~\bibnamefont {Taniguchi}}, \bibinfo {author}
  {\bibfnamefont {K.}~\bibnamefont {Ueno}}, \bibinfo {author} {\bibfnamefont
  {E.}~\bibnamefont {Tutuc}}, \ and\ \bibinfo {author} {\bibfnamefont {B.~J.}\
  \bibnamefont {LeRoy}},\ }\href@noop {} {\bibfield  {journal} {\bibinfo
  {journal} {arXiv preprint arXiv:1910.13068}\ } (\bibinfo {year}
  {2019})}\BibitemShut {NoStop}%
\bibitem [{\citenamefont {Jin}\ \emph {et~al.}(2019)\citenamefont {Jin},
  \citenamefont {Regan}, \citenamefont {Yan}, \citenamefont {Utama},
  \citenamefont {Wang}, \citenamefont {Zhao}, \citenamefont {Qin},
  \citenamefont {Yang}, \citenamefont {Zheng}, \citenamefont {Shi} \emph
  {et~al.}}]{jin2019observation}%
  \BibitemOpen
  \bibfield  {author} {\bibinfo {author} {\bibfnamefont {C.}~\bibnamefont
  {Jin}}, \bibinfo {author} {\bibfnamefont {E.~C.}\ \bibnamefont {Regan}},
  \bibinfo {author} {\bibfnamefont {A.}~\bibnamefont {Yan}}, \bibinfo {author}
  {\bibfnamefont {M.~I.~B.}\ \bibnamefont {Utama}}, \bibinfo {author}
  {\bibfnamefont {D.}~\bibnamefont {Wang}}, \bibinfo {author} {\bibfnamefont
  {S.}~\bibnamefont {Zhao}}, \bibinfo {author} {\bibfnamefont {Y.}~\bibnamefont
  {Qin}}, \bibinfo {author} {\bibfnamefont {S.}~\bibnamefont {Yang}}, \bibinfo
  {author} {\bibfnamefont {Z.}~\bibnamefont {Zheng}}, \bibinfo {author}
  {\bibfnamefont {S.}~\bibnamefont {Shi}},  \emph {et~al.},\ }\href@noop {}
  {\bibfield  {journal} {\bibinfo  {journal} {Nature}\ }\textbf {\bibinfo
  {volume} {567}},\ \bibinfo {pages} {76} (\bibinfo {year} {2019})}\BibitemShut
  {NoStop}%
\bibitem [{\citenamefont {Lindlau}\ \emph {et~al.}(2018)\citenamefont
  {Lindlau}, \citenamefont {Selig}, \citenamefont {Neumann}, \citenamefont
  {Colombier}, \citenamefont {F{\"o}rste}, \citenamefont {Funk}, \citenamefont
  {F{\"o}rg}, \citenamefont {Kim}, \citenamefont {Bergh{\"a}user},
  \citenamefont {Taniguchi} \emph {et~al.}}]{lindlau2018role}%
  \BibitemOpen
  \bibfield  {author} {\bibinfo {author} {\bibfnamefont {J.}~\bibnamefont
  {Lindlau}}, \bibinfo {author} {\bibfnamefont {M.}~\bibnamefont {Selig}},
  \bibinfo {author} {\bibfnamefont {A.}~\bibnamefont {Neumann}}, \bibinfo
  {author} {\bibfnamefont {L.}~\bibnamefont {Colombier}}, \bibinfo {author}
  {\bibfnamefont {J.}~\bibnamefont {F{\"o}rste}}, \bibinfo {author}
  {\bibfnamefont {V.}~\bibnamefont {Funk}}, \bibinfo {author} {\bibfnamefont
  {M.}~\bibnamefont {F{\"o}rg}}, \bibinfo {author} {\bibfnamefont
  {J.}~\bibnamefont {Kim}}, \bibinfo {author} {\bibfnamefont {G.}~\bibnamefont
  {Bergh{\"a}user}}, \bibinfo {author} {\bibfnamefont {T.}~\bibnamefont
  {Taniguchi}},  \emph {et~al.},\ }\href@noop {} {\bibfield  {journal}
  {\bibinfo  {journal} {Nature Communications}\ }\textbf {\bibinfo {volume}
  {9}},\ \bibinfo {pages} {2586} (\bibinfo {year} {2018})}\BibitemShut
  {NoStop}%
\bibitem [{\citenamefont {Deilmann}\ and\ \citenamefont
  {Thygesen}(2019)}]{deilmann2019finite}%
  \BibitemOpen
  \bibfield  {author} {\bibinfo {author} {\bibfnamefont {T.}~\bibnamefont
  {Deilmann}}\ and\ \bibinfo {author} {\bibfnamefont {K.~S.}\ \bibnamefont
  {Thygesen}},\ }\href@noop {} {\bibfield  {journal} {\bibinfo  {journal} {2D
  Materials}\ } (\bibinfo {year} {2019})}\BibitemShut {NoStop}%
\bibitem [{\citenamefont {Bergh{\"a}user}\ \emph {et~al.}(2018)\citenamefont
  {Bergh{\"a}user}, \citenamefont {Steinleitner}, \citenamefont {Merkl},
  \citenamefont {Huber}, \citenamefont {Knorr},\ and\ \citenamefont
  {Malic}}]{berghauser2018mapping}%
  \BibitemOpen
  \bibfield  {author} {\bibinfo {author} {\bibfnamefont {G.}~\bibnamefont
  {Bergh{\"a}user}}, \bibinfo {author} {\bibfnamefont {P.}~\bibnamefont
  {Steinleitner}}, \bibinfo {author} {\bibfnamefont {P.}~\bibnamefont {Merkl}},
  \bibinfo {author} {\bibfnamefont {R.}~\bibnamefont {Huber}}, \bibinfo
  {author} {\bibfnamefont {A.}~\bibnamefont {Knorr}}, \ and\ \bibinfo {author}
  {\bibfnamefont {E.}~\bibnamefont {Malic}},\ }\href@noop {} {\bibfield
  {journal} {\bibinfo  {journal} {Physical Review B}\ }\textbf {\bibinfo
  {volume} {98}},\ \bibinfo {pages} {020301} (\bibinfo {year}
  {2018})}\BibitemShut {NoStop}%
\bibitem [{\citenamefont {Cappelluti}\ \emph {et~al.}(2013)\citenamefont
  {Cappelluti}, \citenamefont {Rold{\'a}n}, \citenamefont {Silva-Guill{\'e}n},
  \citenamefont {Ordej{\'o}n},\ and\ \citenamefont
  {Guinea}}]{cappelluti2013tight}%
  \BibitemOpen
  \bibfield  {author} {\bibinfo {author} {\bibfnamefont {E.}~\bibnamefont
  {Cappelluti}}, \bibinfo {author} {\bibfnamefont {R.}~\bibnamefont
  {Rold{\'a}n}}, \bibinfo {author} {\bibfnamefont {J.}~\bibnamefont
  {Silva-Guill{\'e}n}}, \bibinfo {author} {\bibfnamefont {P.}~\bibnamefont
  {Ordej{\'o}n}}, \ and\ \bibinfo {author} {\bibfnamefont {F.}~\bibnamefont
  {Guinea}},\ }\href@noop {} {\bibfield  {journal} {\bibinfo  {journal}
  {Physical Review B}\ }\textbf {\bibinfo {volume} {88}},\ \bibinfo {pages}
  {075409} (\bibinfo {year} {2013})}\BibitemShut {NoStop}%
\bibitem [{\citenamefont {Rold{\'a}n}\ \emph {et~al.}(2014)\citenamefont
  {Rold{\'a}n}, \citenamefont {Silva-Guill{\'e}n}, \citenamefont
  {L{\'o}pez-Sancho}, \citenamefont {Guinea}, \citenamefont {Cappelluti},\ and\
  \citenamefont {Ordej{\'o}n}}]{roldan2014electronic}%
  \BibitemOpen
  \bibfield  {author} {\bibinfo {author} {\bibfnamefont {R.}~\bibnamefont
  {Rold{\'a}n}}, \bibinfo {author} {\bibfnamefont {J.~A.}\ \bibnamefont
  {Silva-Guill{\'e}n}}, \bibinfo {author} {\bibfnamefont {M.~P.}\ \bibnamefont
  {L{\'o}pez-Sancho}}, \bibinfo {author} {\bibfnamefont {F.}~\bibnamefont
  {Guinea}}, \bibinfo {author} {\bibfnamefont {E.}~\bibnamefont {Cappelluti}},
  \ and\ \bibinfo {author} {\bibfnamefont {P.}~\bibnamefont {Ordej{\'o}n}},\
  }\href@noop {} {\bibfield  {journal} {\bibinfo  {journal} {Annalen der
  Physik}\ }\textbf {\bibinfo {volume} {526}},\ \bibinfo {pages} {347}
  (\bibinfo {year} {2014})}\BibitemShut {NoStop}%
\bibitem [{\citenamefont {He}\ \emph {et~al.}(2014)\citenamefont {He},
  \citenamefont {Kumar}, \citenamefont {Zhao}, \citenamefont {Wang},
  \citenamefont {Mak}, \citenamefont {Zhao},\ and\ \citenamefont
  {Shan}}]{he2014tightly}%
  \BibitemOpen
  \bibfield  {author} {\bibinfo {author} {\bibfnamefont {K.}~\bibnamefont
  {He}}, \bibinfo {author} {\bibfnamefont {N.}~\bibnamefont {Kumar}}, \bibinfo
  {author} {\bibfnamefont {L.}~\bibnamefont {Zhao}}, \bibinfo {author}
  {\bibfnamefont {Z.}~\bibnamefont {Wang}}, \bibinfo {author} {\bibfnamefont
  {K.~F.}\ \bibnamefont {Mak}}, \bibinfo {author} {\bibfnamefont
  {H.}~\bibnamefont {Zhao}}, \ and\ \bibinfo {author} {\bibfnamefont
  {J.}~\bibnamefont {Shan}},\ }\href@noop {} {\bibfield  {journal} {\bibinfo
  {journal} {Physical review letters}\ }\textbf {\bibinfo {volume} {113}},\
  \bibinfo {pages} {026803} (\bibinfo {year} {2014})}\BibitemShut {NoStop}%
\bibitem [{\citenamefont {Wang}\ \emph {et~al.}(2018)\citenamefont {Wang},
  \citenamefont {Chernikov}, \citenamefont {Glazov}, \citenamefont {Heinz},
  \citenamefont {Marie}, \citenamefont {Amand},\ and\ \citenamefont
  {Urbaszek}}]{wang2018colloquium}%
  \BibitemOpen
  \bibfield  {author} {\bibinfo {author} {\bibfnamefont {G.}~\bibnamefont
  {Wang}}, \bibinfo {author} {\bibfnamefont {A.}~\bibnamefont {Chernikov}},
  \bibinfo {author} {\bibfnamefont {M.~M.}\ \bibnamefont {Glazov}}, \bibinfo
  {author} {\bibfnamefont {T.~F.}\ \bibnamefont {Heinz}}, \bibinfo {author}
  {\bibfnamefont {X.}~\bibnamefont {Marie}}, \bibinfo {author} {\bibfnamefont
  {T.}~\bibnamefont {Amand}}, \ and\ \bibinfo {author} {\bibfnamefont
  {B.}~\bibnamefont {Urbaszek}},\ }\href@noop {} {\bibfield  {journal}
  {\bibinfo  {journal} {Reviews of Modern Physics}\ }\textbf {\bibinfo {volume}
  {90}},\ \bibinfo {pages} {021001} (\bibinfo {year} {2018})}\BibitemShut
  {NoStop}%
\bibitem [{\citenamefont {Mueller}\ and\ \citenamefont
  {Malic}(2018)}]{mueller2018exciton}%
  \BibitemOpen
  \bibfield  {author} {\bibinfo {author} {\bibfnamefont {T.}~\bibnamefont
  {Mueller}}\ and\ \bibinfo {author} {\bibfnamefont {E.}~\bibnamefont
  {Malic}},\ }\href@noop {} {\bibfield  {journal} {\bibinfo  {journal} {npj 2D
  Materials and Applications}\ }\textbf {\bibinfo {volume} {2}},\ \bibinfo
  {pages} {1} (\bibinfo {year} {2018})}\BibitemShut {NoStop}%
\bibitem [{\citenamefont {Ovesen}\ \emph {et~al.}(2019)\citenamefont {Ovesen},
  \citenamefont {Brem}, \citenamefont {Linder{\"a}lv}, \citenamefont {Kuisma},
  \citenamefont {Korn}, \citenamefont {Erhart}, \citenamefont {Selig},\ and\
  \citenamefont {Malic}}]{ovesen2019interlayer}%
  \BibitemOpen
  \bibfield  {author} {\bibinfo {author} {\bibfnamefont {S.}~\bibnamefont
  {Ovesen}}, \bibinfo {author} {\bibfnamefont {S.}~\bibnamefont {Brem}},
  \bibinfo {author} {\bibfnamefont {C.}~\bibnamefont {Linder{\"a}lv}}, \bibinfo
  {author} {\bibfnamefont {M.}~\bibnamefont {Kuisma}}, \bibinfo {author}
  {\bibfnamefont {T.}~\bibnamefont {Korn}}, \bibinfo {author} {\bibfnamefont
  {P.}~\bibnamefont {Erhart}}, \bibinfo {author} {\bibfnamefont
  {M.}~\bibnamefont {Selig}}, \ and\ \bibinfo {author} {\bibfnamefont
  {E.}~\bibnamefont {Malic}},\ }\href@noop {} {\bibfield  {journal} {\bibinfo
  {journal} {Communications Physics}\ }\textbf {\bibinfo {volume} {2}},\
  \bibinfo {pages} {1} (\bibinfo {year} {2019})}\BibitemShut {NoStop}%
\bibitem [{\citenamefont {Ruiz-Tijerina}\ and\ \citenamefont
  {Fal'ko}(2019)}]{ruiz2019interlayer}%
  \BibitemOpen
  \bibfield  {author} {\bibinfo {author} {\bibfnamefont {D.~A.}\ \bibnamefont
  {Ruiz-Tijerina}}\ and\ \bibinfo {author} {\bibfnamefont {V.~I.}\ \bibnamefont
  {Fal'ko}},\ }\href@noop {} {\bibfield  {journal} {\bibinfo  {journal}
  {Physical Review B}\ }\textbf {\bibinfo {volume} {99}},\ \bibinfo {pages}
  {125424} (\bibinfo {year} {2019})}\BibitemShut {NoStop}%
\bibitem [{\citenamefont {Wang}\ \emph {et~al.}(2017)\citenamefont {Wang},
  \citenamefont {Wang}, \citenamefont {Yao}, \citenamefont {Liu},\ and\
  \citenamefont {Yu}}]{wang2017interlayer}%
  \BibitemOpen
  \bibfield  {author} {\bibinfo {author} {\bibfnamefont {Y.}~\bibnamefont
  {Wang}}, \bibinfo {author} {\bibfnamefont {Z.}~\bibnamefont {Wang}}, \bibinfo
  {author} {\bibfnamefont {W.}~\bibnamefont {Yao}}, \bibinfo {author}
  {\bibfnamefont {G.-B.}\ \bibnamefont {Liu}}, \ and\ \bibinfo {author}
  {\bibfnamefont {H.}~\bibnamefont {Yu}},\ }\href@noop {} {\bibfield  {journal}
  {\bibinfo  {journal} {Physical Review B}\ }\textbf {\bibinfo {volume} {95}},\
  \bibinfo {pages} {115429} (\bibinfo {year} {2017})}\BibitemShut {NoStop}%
\bibitem [{\citenamefont {Korm{\'a}nyos}\ \emph {et~al.}(2015)\citenamefont
  {Korm{\'a}nyos}, \citenamefont {Burkard}, \citenamefont {Gmitra},
  \citenamefont {Fabian}, \citenamefont {Z{\'o}lyomi}, \citenamefont
  {Drummond},\ and\ \citenamefont {Falko}}]{kormanyos2015k}%
  \BibitemOpen
  \bibfield  {author} {\bibinfo {author} {\bibfnamefont {A.}~\bibnamefont
  {Korm{\'a}nyos}}, \bibinfo {author} {\bibfnamefont {G.}~\bibnamefont
  {Burkard}}, \bibinfo {author} {\bibfnamefont {M.}~\bibnamefont {Gmitra}},
  \bibinfo {author} {\bibfnamefont {J.}~\bibnamefont {Fabian}}, \bibinfo
  {author} {\bibfnamefont {V.}~\bibnamefont {Z{\'o}lyomi}}, \bibinfo {author}
  {\bibfnamefont {N.~D.}\ \bibnamefont {Drummond}}, \ and\ \bibinfo {author}
  {\bibfnamefont {V.}~\bibnamefont {Falko}},\ }\href@noop {} {\bibfield
  {journal} {\bibinfo  {journal} {2D Materials}\ }\textbf {\bibinfo {volume}
  {2}},\ \bibinfo {pages} {022001} (\bibinfo {year} {2015})}\BibitemShut
  {NoStop}%
\bibitem [{\citenamefont {Katsch}\ \emph {et~al.}(2018)\citenamefont {Katsch},
  \citenamefont {Selig}, \citenamefont {Carmele},\ and\ \citenamefont
  {Knorr}}]{katsch2018theory}%
  \BibitemOpen
  \bibfield  {author} {\bibinfo {author} {\bibfnamefont {F.}~\bibnamefont
  {Katsch}}, \bibinfo {author} {\bibfnamefont {M.}~\bibnamefont {Selig}},
  \bibinfo {author} {\bibfnamefont {A.}~\bibnamefont {Carmele}}, \ and\
  \bibinfo {author} {\bibfnamefont {A.}~\bibnamefont {Knorr}},\ }\href@noop {}
  {\bibfield  {journal} {\bibinfo  {journal} {Physica Status Solidi (b)}\
  }\textbf {\bibinfo {volume} {255}},\ \bibinfo {pages} {1800185} (\bibinfo
  {year} {2018})}\BibitemShut {NoStop}%
\bibitem [{\citenamefont {Ivanov}\ and\ \citenamefont
  {Haug}(1993)}]{ivanov1993self}%
  \BibitemOpen
  \bibfield  {author} {\bibinfo {author} {\bibfnamefont {A.}~\bibnamefont
  {Ivanov}}\ and\ \bibinfo {author} {\bibfnamefont {H.}~\bibnamefont {Haug}},\
  }\href@noop {} {\bibfield  {journal} {\bibinfo  {journal} {Physical Review
  B}\ }\textbf {\bibinfo {volume} {48}},\ \bibinfo {pages} {1490} (\bibinfo
  {year} {1993})}\BibitemShut {NoStop}%
\bibitem [{\citenamefont {Toyozawa}(1958)}]{toyozawa1958theory}%
  \BibitemOpen
  \bibfield  {author} {\bibinfo {author} {\bibfnamefont {Y.}~\bibnamefont
  {Toyozawa}},\ }\href@noop {} {\bibfield  {journal} {\bibinfo  {journal}
  {Progress of Theoretical Physics}\ }\textbf {\bibinfo {volume} {20}},\
  \bibinfo {pages} {53} (\bibinfo {year} {1958})}\BibitemShut {NoStop}%
\bibitem [{\citenamefont {Laturia}\ \emph {et~al.}(2018)\citenamefont
  {Laturia}, \citenamefont {Van~de Put},\ and\ \citenamefont
  {Vandenberghe}}]{laturia2018dielectric}%
  \BibitemOpen
  \bibfield  {author} {\bibinfo {author} {\bibfnamefont {A.}~\bibnamefont
  {Laturia}}, \bibinfo {author} {\bibfnamefont {M.~L.}\ \bibnamefont {Van~de
  Put}}, \ and\ \bibinfo {author} {\bibfnamefont {W.~G.}\ \bibnamefont
  {Vandenberghe}},\ }\href@noop {} {\bibfield  {journal} {\bibinfo  {journal}
  {NPJ 2D Materials and Applications}\ }\textbf {\bibinfo {volume} {2}},\
  \bibinfo {pages} {6} (\bibinfo {year} {2018})}\BibitemShut {NoStop}%
\bibitem [{\citenamefont {Gillen}\ and\ \citenamefont
  {Maultzsch}(2018)}]{gillen2018interlayer}%
  \BibitemOpen
  \bibfield  {author} {\bibinfo {author} {\bibfnamefont {R.}~\bibnamefont
  {Gillen}}\ and\ \bibinfo {author} {\bibfnamefont {J.}~\bibnamefont
  {Maultzsch}},\ }\href@noop {} {\bibfield  {journal} {\bibinfo  {journal}
  {Physical Review B}\ }\textbf {\bibinfo {volume} {97}},\ \bibinfo {pages}
  {165306} (\bibinfo {year} {2018})}\BibitemShut {NoStop}%
\bibitem [{\citenamefont {Zheng}\ \emph {et~al.}(2014)\citenamefont {Zheng},
  \citenamefont {Feng},\ and\ \citenamefont {Yong-Shi}}]{zheng2014exotic}%
  \BibitemOpen
  \bibfield  {author} {\bibinfo {author} {\bibfnamefont {L.}~\bibnamefont
  {Zheng}}, \bibinfo {author} {\bibfnamefont {L.}~\bibnamefont {Feng}}, \ and\
  \bibinfo {author} {\bibfnamefont {W.}~\bibnamefont {Yong-Shi}},\ }\href@noop
  {} {\bibfield  {journal} {\bibinfo  {journal} {Chinese Physics B}\ }\textbf
  {\bibinfo {volume} {23}},\ \bibinfo {pages} {077308} (\bibinfo {year}
  {2014})}\BibitemShut {NoStop}%
\bibitem [{\citenamefont {Selig}\ \emph {et~al.}(2016)\citenamefont {Selig},
  \citenamefont {Bergh{\"a}user}, \citenamefont {Raja}, \citenamefont {Nagler},
  \citenamefont {Sch{\"u}ller}, \citenamefont {Heinz}, \citenamefont {Korn},
  \citenamefont {Chernikov}, \citenamefont {Malic},\ and\ \citenamefont
  {Knorr}}]{selig2016excitonic}%
  \BibitemOpen
  \bibfield  {author} {\bibinfo {author} {\bibfnamefont {M.}~\bibnamefont
  {Selig}}, \bibinfo {author} {\bibfnamefont {G.}~\bibnamefont
  {Bergh{\"a}user}}, \bibinfo {author} {\bibfnamefont {A.}~\bibnamefont
  {Raja}}, \bibinfo {author} {\bibfnamefont {P.}~\bibnamefont {Nagler}},
  \bibinfo {author} {\bibfnamefont {C.}~\bibnamefont {Sch{\"u}ller}}, \bibinfo
  {author} {\bibfnamefont {T.~F.}\ \bibnamefont {Heinz}}, \bibinfo {author}
  {\bibfnamefont {T.}~\bibnamefont {Korn}}, \bibinfo {author} {\bibfnamefont
  {A.}~\bibnamefont {Chernikov}}, \bibinfo {author} {\bibfnamefont
  {E.}~\bibnamefont {Malic}}, \ and\ \bibinfo {author} {\bibfnamefont
  {A.}~\bibnamefont {Knorr}},\ }\href@noop {} {\bibfield  {journal} {\bibinfo
  {journal} {Nature Communications}\ }\textbf {\bibinfo {volume} {7}},\
  \bibinfo {pages} {13279} (\bibinfo {year} {2016})}\BibitemShut {NoStop}%
\bibitem [{\citenamefont {Zhang}\ \emph {et~al.}(2015)\citenamefont {Zhang},
  \citenamefont {You}, \citenamefont {Zhao},\ and\ \citenamefont
  {Heinz}}]{zhang2015experimental}%
  \BibitemOpen
  \bibfield  {author} {\bibinfo {author} {\bibfnamefont {X.-X.}\ \bibnamefont
  {Zhang}}, \bibinfo {author} {\bibfnamefont {Y.}~\bibnamefont {You}}, \bibinfo
  {author} {\bibfnamefont {S.~Y.~F.}\ \bibnamefont {Zhao}}, \ and\ \bibinfo
  {author} {\bibfnamefont {T.~F.}\ \bibnamefont {Heinz}},\ }\href@noop {}
  {\bibfield  {journal} {\bibinfo  {journal} {Physical Review Letters}\
  }\textbf {\bibinfo {volume} {115}},\ \bibinfo {pages} {257403} (\bibinfo
  {year} {2015})}\BibitemShut {NoStop}%
\bibitem [{\citenamefont {Niehues}\ \emph {et~al.}(2018)\citenamefont
  {Niehues}, \citenamefont {Schmidt}, \citenamefont {Drüppel}, \citenamefont
  {Marauhn}, \citenamefont {Christiansen}, \citenamefont {Selig}, \citenamefont
  {Berghäuser}, \citenamefont {Wigger}, \citenamefont {Schneider},
  \citenamefont {Braasch} \emph {et~al.}}]{niehues2018strain}%
  \BibitemOpen
  \bibfield  {author} {\bibinfo {author} {\bibfnamefont {I.}~\bibnamefont
  {Niehues}}, \bibinfo {author} {\bibfnamefont {R.}~\bibnamefont {Schmidt}},
  \bibinfo {author} {\bibfnamefont {M.}~\bibnamefont {Drüppel}}, \bibinfo
  {author} {\bibfnamefont {P.}~\bibnamefont {Marauhn}}, \bibinfo {author}
  {\bibfnamefont {D.}~\bibnamefont {Christiansen}}, \bibinfo {author}
  {\bibfnamefont {M.}~\bibnamefont {Selig}}, \bibinfo {author} {\bibfnamefont
  {G.}~\bibnamefont {Berghäuser}}, \bibinfo {author} {\bibfnamefont
  {D.}~\bibnamefont {Wigger}}, \bibinfo {author} {\bibfnamefont
  {R.}~\bibnamefont {Schneider}}, \bibinfo {author} {\bibfnamefont
  {L.}~\bibnamefont {Braasch}},  \emph {et~al.},\ }\href@noop {} {\bibfield
  {journal} {\bibinfo  {journal} {Nano Letters}\ }\textbf {\bibinfo {volume}
  {18}},\ \bibinfo {pages} {1751} (\bibinfo {year} {2018})}\BibitemShut
  {NoStop}%
\bibitem [{\citenamefont {Ye}\ \emph {et~al.}(2016)\citenamefont {Ye},
  \citenamefont {Dou}, \citenamefont {Ding}, \citenamefont {Jiang},
  \citenamefont {Yang},\ and\ \citenamefont {Sun}}]{ye2016pressure}%
  \BibitemOpen
  \bibfield  {author} {\bibinfo {author} {\bibfnamefont {Y.}~\bibnamefont
  {Ye}}, \bibinfo {author} {\bibfnamefont {X.}~\bibnamefont {Dou}}, \bibinfo
  {author} {\bibfnamefont {K.}~\bibnamefont {Ding}}, \bibinfo {author}
  {\bibfnamefont {D.}~\bibnamefont {Jiang}}, \bibinfo {author} {\bibfnamefont
  {F.}~\bibnamefont {Yang}}, \ and\ \bibinfo {author} {\bibfnamefont
  {B.}~\bibnamefont {Sun}},\ }\href@noop {} {\bibfield  {journal} {\bibinfo
  {journal} {Nanoscale}\ }\textbf {\bibinfo {volume} {8}},\ \bibinfo {pages}
  {10843} (\bibinfo {year} {2016})}\BibitemShut {NoStop}%
\bibitem [{\citenamefont {Brem}\ \emph {et~al.}(2020)\citenamefont {Brem},
  \citenamefont {Ekman}, \citenamefont {Christiansen}, \citenamefont {Katsch},
  \citenamefont {Selig}, \citenamefont {Robert}, \citenamefont {Marie},
  \citenamefont {Urbaszek}, \citenamefont {Knorr},\ and\ \citenamefont
  {Malic}}]{brem2020phonon}%
  \BibitemOpen
  \bibfield  {author} {\bibinfo {author} {\bibfnamefont {S.}~\bibnamefont
  {Brem}}, \bibinfo {author} {\bibfnamefont {A.}~\bibnamefont {Ekman}},
  \bibinfo {author} {\bibfnamefont {D.}~\bibnamefont {Christiansen}}, \bibinfo
  {author} {\bibfnamefont {F.}~\bibnamefont {Katsch}}, \bibinfo {author}
  {\bibfnamefont {M.}~\bibnamefont {Selig}}, \bibinfo {author} {\bibfnamefont
  {C.}~\bibnamefont {Robert}}, \bibinfo {author} {\bibfnamefont
  {X.}~\bibnamefont {Marie}}, \bibinfo {author} {\bibfnamefont
  {B.}~\bibnamefont {Urbaszek}}, \bibinfo {author} {\bibfnamefont
  {A.}~\bibnamefont {Knorr}}, \ and\ \bibinfo {author} {\bibfnamefont
  {E.}~\bibnamefont {Malic}},\ }\href@noop {} {\bibfield  {journal} {\bibinfo
  {journal} {Nano Letters}\ } (\bibinfo {year} {2020})}\BibitemShut {NoStop}%
\bibitem [{\citenamefont {Kira}\ and\ \citenamefont
  {Koch}(2011)}]{kira2011semiconductor}%
  \BibitemOpen
  \bibfield  {author} {\bibinfo {author} {\bibfnamefont {M.}~\bibnamefont
  {Kira}}\ and\ \bibinfo {author} {\bibfnamefont {S.~W.}\ \bibnamefont
  {Koch}},\ }\href@noop {} {\emph {\bibinfo {title} {Semiconductor quantum
  optics}}}\ (\bibinfo  {publisher} {Cambridge University Press},\ \bibinfo
  {year} {2011})\BibitemShut {NoStop}%
\bibitem [{\citenamefont {Wu}\ \emph {et~al.}(2017)\citenamefont {Wu},
  \citenamefont {Lovorn},\ and\ \citenamefont {MacDonald}}]{wu2017topological}%
  \BibitemOpen
  \bibfield  {author} {\bibinfo {author} {\bibfnamefont {F.}~\bibnamefont
  {Wu}}, \bibinfo {author} {\bibfnamefont {T.}~\bibnamefont {Lovorn}}, \ and\
  \bibinfo {author} {\bibfnamefont {A.~H.}\ \bibnamefont {MacDonald}},\
  }\href@noop {} {\bibfield  {journal} {\bibinfo  {journal} {Physical review
  letters}\ }\textbf {\bibinfo {volume} {118}},\ \bibinfo {pages} {147401}
  (\bibinfo {year} {2017})}\BibitemShut {NoStop}%
\bibitem [{\citenamefont {Jin}\ \emph {et~al.}(2014)\citenamefont {Jin},
  \citenamefont {Li}, \citenamefont {Mullen},\ and\ \citenamefont
  {Kim}}]{jin2014intrinsic}%
  \BibitemOpen
  \bibfield  {author} {\bibinfo {author} {\bibfnamefont {Z.}~\bibnamefont
  {Jin}}, \bibinfo {author} {\bibfnamefont {X.}~\bibnamefont {Li}}, \bibinfo
  {author} {\bibfnamefont {J.~T.}\ \bibnamefont {Mullen}}, \ and\ \bibinfo
  {author} {\bibfnamefont {K.~W.}\ \bibnamefont {Kim}},\ }\href@noop {}
  {\bibfield  {journal} {\bibinfo  {journal} {Physical Review B}\ }\textbf
  {\bibinfo {volume} {90}},\ \bibinfo {pages} {045422} (\bibinfo {year}
  {2014})}\BibitemShut {NoStop}%
\bibitem [{\citenamefont {Selig}\ \emph {et~al.}(2018)\citenamefont {Selig},
  \citenamefont {Bergh{\"a}user}, \citenamefont {Richter}, \citenamefont
  {Bratschitsch}, \citenamefont {Knorr},\ and\ \citenamefont
  {Malic}}]{selig2018dark}%
  \BibitemOpen
  \bibfield  {author} {\bibinfo {author} {\bibfnamefont {M.}~\bibnamefont
  {Selig}}, \bibinfo {author} {\bibfnamefont {G.}~\bibnamefont
  {Bergh{\"a}user}}, \bibinfo {author} {\bibfnamefont {M.}~\bibnamefont
  {Richter}}, \bibinfo {author} {\bibfnamefont {R.}~\bibnamefont
  {Bratschitsch}}, \bibinfo {author} {\bibfnamefont {A.}~\bibnamefont {Knorr}},
  \ and\ \bibinfo {author} {\bibfnamefont {E.}~\bibnamefont {Malic}},\
  }\href@noop {} {\bibfield  {journal} {\bibinfo  {journal} {2D Materials}\
  }\textbf {\bibinfo {volume} {5}},\ \bibinfo {pages} {035017} (\bibinfo {year}
  {2018})}\BibitemShut {NoStop}%
\bibitem [{\citenamefont {Brem}\ \emph {et~al.}(2018)\citenamefont {Brem},
  \citenamefont {Selig}, \citenamefont {Berghaeuser},\ and\ \citenamefont
  {Malic}}]{brem2018exciton}%
  \BibitemOpen
  \bibfield  {author} {\bibinfo {author} {\bibfnamefont {S.}~\bibnamefont
  {Brem}}, \bibinfo {author} {\bibfnamefont {M.}~\bibnamefont {Selig}},
  \bibinfo {author} {\bibfnamefont {G.}~\bibnamefont {Berghaeuser}}, \ and\
  \bibinfo {author} {\bibfnamefont {E.}~\bibnamefont {Malic}},\ }\href@noop {}
  {\bibfield  {journal} {\bibinfo  {journal} {Scientific Reports}\ }\textbf
  {\bibinfo {volume} {8}},\ \bibinfo {pages} {8238} (\bibinfo {year}
  {2018})}\BibitemShut {NoStop}%
\bibitem [{\citenamefont {Lin}\ \emph {et~al.}(2018)\citenamefont {Lin},
  \citenamefont {Tan}, \citenamefont {Wu}, \citenamefont {Chen}, \citenamefont
  {Wang}, \citenamefont {Pan}, \citenamefont {Zhang}, \citenamefont {Cong},
  \citenamefont {Zhang}, \citenamefont {Ji} \emph {et~al.}}]{lin2018moire}%
  \BibitemOpen
  \bibfield  {author} {\bibinfo {author} {\bibfnamefont {M.-L.}\ \bibnamefont
  {Lin}}, \bibinfo {author} {\bibfnamefont {Q.-H.}\ \bibnamefont {Tan}},
  \bibinfo {author} {\bibfnamefont {J.-B.}\ \bibnamefont {Wu}}, \bibinfo
  {author} {\bibfnamefont {X.-S.}\ \bibnamefont {Chen}}, \bibinfo {author}
  {\bibfnamefont {J.-H.}\ \bibnamefont {Wang}}, \bibinfo {author}
  {\bibfnamefont {Y.-H.}\ \bibnamefont {Pan}}, \bibinfo {author} {\bibfnamefont
  {X.}~\bibnamefont {Zhang}}, \bibinfo {author} {\bibfnamefont
  {X.}~\bibnamefont {Cong}}, \bibinfo {author} {\bibfnamefont {J.}~\bibnamefont
  {Zhang}}, \bibinfo {author} {\bibfnamefont {W.}~\bibnamefont {Ji}},  \emph
  {et~al.},\ }\href@noop {} {\bibfield  {journal} {\bibinfo  {journal} {ACS
  nano}\ }\textbf {\bibinfo {volume} {12}},\ \bibinfo {pages} {8770} (\bibinfo
  {year} {2018})}\BibitemShut {NoStop}%
\end{thebibliography}
\end{document}